\documentclass[%
 reprint,
 amsmath,amssymb,
 aps,
pre,
]{revtex4-1}
\pdfoutput=1

\usepackage{hyperref}

\usepackage{subfigure}
\usepackage[pdftex]{graphicx}

\hypersetup{ colorlinks,linkcolor=blue,filecolor=blue,urlcolor=blue,citecolor=blue }

\begin{document}


\title{Phase diagrams of the ZGB model on random networks}

\author{E. B. Vilela}
\author{H. A. Fernandes}
\author{F. L. P. Costa}
\author{P. F. Gomes}
\email{paulofreitasgomes@ufg.br}
\affiliation{%
Instituto de Ci\^encias Exatas, Universidade Federal de Jata\'i, BR 364, km 195, 75801-615, Jata\'i-GO, Brazil.
}%


\date{\today}

\begin{abstract}
In this work, we revisited the ZGB model in order to study the behavior of its phase diagram when two well-known random networks play the role of the catalytic surfaces: the Random Geometric Graph and the Erd\"{o}s-R\'{e}nyi network. The connectivity and, therefore, the average number of neighbors of the nodes of these networks can vary according to their control parameters, the neighborhood radius $\alpha$ and the linking probability $p$, respectively. In addition, the catalytic reactions of the ZGB model are governed by the parameter $y$, the adsorption rate of carbon monoxide molecules on the catalytic surface. So, to study the phase diagrams of the model on both random networks, we carried out extensive steady-state Monte Carlo simulations in the space parameters ($y,\alpha$) and ($y,p$) and showed that the continuous phase transition is greatly affected by the topological features of the networks while the discontinuous one remains present in the diagram throughout the interval of study.
\end{abstract}

\pacs{Valid PACS appear here}
\maketitle


\section{Introduction}

Phase transitions and critical phenomena of equilibrium systems have, for a long time, been studied mainly through Monte Carlo simulations, making it one of the most important methods in Statistical Mechanics. More recently, nonequilibrium systems have also attracted a lot of attention, since it was discovered that they can exhibit similar behavior to those found in equilibrium systems. One of those systems which has been extensively studied and has become a prototype in nonequilibrium Monte Carlo simulations is the surface reaction model known as ZGB model. In their pioneering work, R.M. Ziff, E. Gulari, and Y. Barshad \cite{ziff1986} proposed a model that describes some nonequilibrium aspects of the production of carbon dioxide ($\textrm{CO}_2$) molecules through the reaction of carbon monoxide ($\textrm{CO}$) molecules with oxygen (O) atoms, both adsorbed on a catalytic surface represented by square lattices. Despite its simplicity, this model presents a rich phase diagram with two irreversible phase transitions, one continuous and another discontinuous, separating the reactive phase from two absorbing phases \cite{ziff1986,meakin1987,fisher1989}. In addition, this model had shown to be capable of reproducing some aspects of transition-metal catalysts being, therefore, of interest for possible technological applications \cite{bond1987, zhdanov1994, marro1999}. For instance, discontinuous phase transitions are observed in some experimental works on platinum \cite{golchet1978, matsushima1979, ehsasi1989, christmann1991, block1993, berdau1999} even though there are no experimental evidences of the continuous phase transition presented in the ZGB model. In fact, reaction processes in catalytic surfaces have also gained much attention due to their importance on the petrochemical and automotive industries \cite{Hagen2006}. 

Since its discovery, several modified versions of the original model have been proposed. Similarly, some works have considered the desorption and/or diffusion \cite{fischer1989, dumont1990, albano1992, brosilow1992, tome1993, kaukonen1989, jensen1990, buendia2009, grandi2002, buendia2013, chan2015, buendia2015, dasilva2018, fernandes2018} of CO molecules. Other works included the presence of impurities on the surface \cite{hoenicke2000, buendia2012, buendia2013,buendia2015,hoenicke2014,fernandes2019} or took into account the attractive and/or repulsive interactions between the adsorbed molecules \cite{buendia2015,satulovsky1992}. Notably, most numerical works on the ZGB model has considered regular square lattices (SL) as the catalytic surfaces where the reactions occur. On the other hand, only few works took into account different network structures, such as hexagonal \cite{meakin1987, provata2005, noussiou2007, provata2007}, fractal \cite{albano1990,mai1992,albano1994,gao1999a,gao1999b} and Voronoi-Delaunay random lattices \cite{Oliveira2016}. Khan and Yaldram \cite{khan2000} have also considered a two-layer structure built through the interaction among the sites of two square lattices. A simple way to create a flexibility on the pattern of the sites communication is to arrange them in a more sophisticated manner and a natural suggestion is a simple random network. Unlike regular networks, e.g. square or hexagonal lattices, in which there exists only a single node degree (each site, or node, has a fix number of connections or neighbors), the random networks present a tractable degree distribution thus creating a whole new set of topological properties.


In this work, we have revisited the ZGB model in order to study the behavior of its phase transitions through extensive steady-state Monte Carlo simulations on two different and well-known random networks: the Random Geometric Graph (RGG) and the Erd\"{o}s-R\'{e}nyi network (ERN). These two networks are used as prototype for communication patterns and the connections between their sites follow precise rules defined by their control parameters.

This paper is organized as follows: in Sec. \ref{model} we describe the ZGB model and its phase diagram, as well as the two random networks considered in this study. In Sec. \ref{results} we show our main results and present the discussions about our findings. Finally, the conclusions are presented in Sec. \ref{sec:conclusions}. 

\section{The model} \label{model}

\subsection{The Ziff-Gulari-Barshad model}

The Ziff-Gulari-Barshad (ZGB) model \cite{ziff1986} is a catalytic surface model that mimics the production of carbon dioxide (CO$_2$) molecules when carbon monoxide (CO) and oxygen (O$_2$) molecules, both in the gas phase, impinge the surface with rates $y$ and $1-y$, respectively. In the adsorption process, the O$_2$ molecules dissociate into two oxygen (O) atoms and, therefore, the sites of the surface can be filled with O atoms, CO molecules, or be vacant ($V$). Once adsorbed on the surface, the CO molecules and O atoms can react producing CO$_2$ molecules. The set of reactions follows the Langmuir-Hinshelwood mechanism \cite{evans1991} and can be represented by the following reaction equations: 
\begin{align}
\textrm{CO}(g)+V &\longrightarrow \textrm{CO}(a), \label{reaction1} \\
\textrm{O}_2(g)+2V &\longrightarrow 2\textrm{O}(a), \label{reaction2} \\
\textrm{CO}(a)+\textrm{O}(a) &\longrightarrow \textrm{CO}_2(g)+2V.  \label{eq:reaction}
\end{align}
Eq. \ref{reaction1} is related to the CO molecule adsorption process. The molecule in the gas ($g$) phase impinges the surface with a rate $y$ being adsorbed ($a$) on it only if the randomly chosen site is vacant. On the other hand, as stated by Eq. \ref{reaction2}, if the O$_2$ molecule is chosen (with a rate $1-y$), it dissociates into two O atoms and both are adsorbed on the surface only if the two neighbor sites, also chosen at random, are empty. If any of the adsorption sites is occupied, the adsorption processes do not occur and the O$_2$ molecule returns to the gas phase. The production of CO$_2$ molecules, which is represented by Eq. \ref{eq:reaction}, occurs whenever O atoms and CO molecules are neighbors on the surface. In this reaction, the CO$_2$ molecule desorbs from the surface leaving two vacant sites behind. As shown above, $y$ is the sole parameter that controls the kinetics of the ZGB model. 

Regardless its simplicity, the model presents two irreversible phase transitions when implemented on the regular square lattice (SL). The first one is a continuous phase transition which takes place at $y_1 \simeq 0.3874$ \cite{voigt1997} and separates the absorbing phase where system is poisoned by O atoms ($0 \leq y < y_1$) and the active phase ($y_1 \leq y < y_2$) in which there exists the production of CO$_2$ molecules. This active phase persists until $y_2 \simeq 0.5256$ \cite{ziff1992} when a discontinuous phase transition occurs and the system is suddenly trapped in the CO absorbing state ($y_2 < y \leq 1$). 

It is notable that this simple model is capable to describe the discontinuous phase transition experimentally verified by several authors, as stated above. In contempt of its existence in theoretical studies, there are no experimental evidences of the continuous phase transition. This last transition has also being studied by several authors with results supporting that the critical point belongs to the directed percolation (DP) universality class \cite{jensen1990a, grinstein1989,Fernandes2016}.


\subsection{Random Networks}

In this work, we study the ZGB model in which the catalytic surface is represented by two possible random networks: the Random Geometric Graph \cite{penrose2003,spodarev2013} and the Erd\"{o}s-R\'{e}nyi network \cite{erdos1959,erdos1960,Gilbert1959}. A network (or a graph) is a set of vertices (or nodes) connected by a list of links (or edges) \cite{Newman2010,Barabasi2016}. Here, the vertices are the sites of the catalytic surfaces and the links representing the connectivity among the vertices provide the neighbors of each site. Each node and its connectivity are randomly defined on the catalytic surface according to the criterion for each considered random network. This criterion creates a distribution of all network properties such as degree (number of neighbors of each node), centrality, number of components and others. On the contrary, for the regular square and hexagonal lattices, each site has exactly four and six neighbors, respectively. Some of the questions we will try to answer in this work are the following: Are the continuous and discontinuous phase transitions present on both random networks? What is the effect of the connectivity on the properties of the ZGB model? Before answering these questions, we recapitulate the main properties of the random networks considered in this study and compare them to the square lattice (SL).

\subsubsection{The Random Geometric Graph (RGG)}

The RGG is constructed by placing $M$ nodes (sites) at random in a square box of linear size $L$. We maintain the superficial density of sites $d = M/L^2$ constant in order to discard its effect when the size $M$ of the network is adjusted. In this manner, for a given $M$ we set $L = \sqrt{M/ d}$. Since the effective area of one site is $1/d$, we adopt the linear size $\ell = 1/\sqrt{d}$ as the spatial scale to measure all distances in our study \cite{Gomes2019}. The neighborhood of a given site, as well as the network properties, is defined by a single control parameter $r = \alpha \ell$ with $\alpha>0$, the neighborhood radius. Figure 1 shows a snapshot of a network with $M=256$.
\begin{figure}[h!]
\centering
\includegraphics[width=3.0 in]{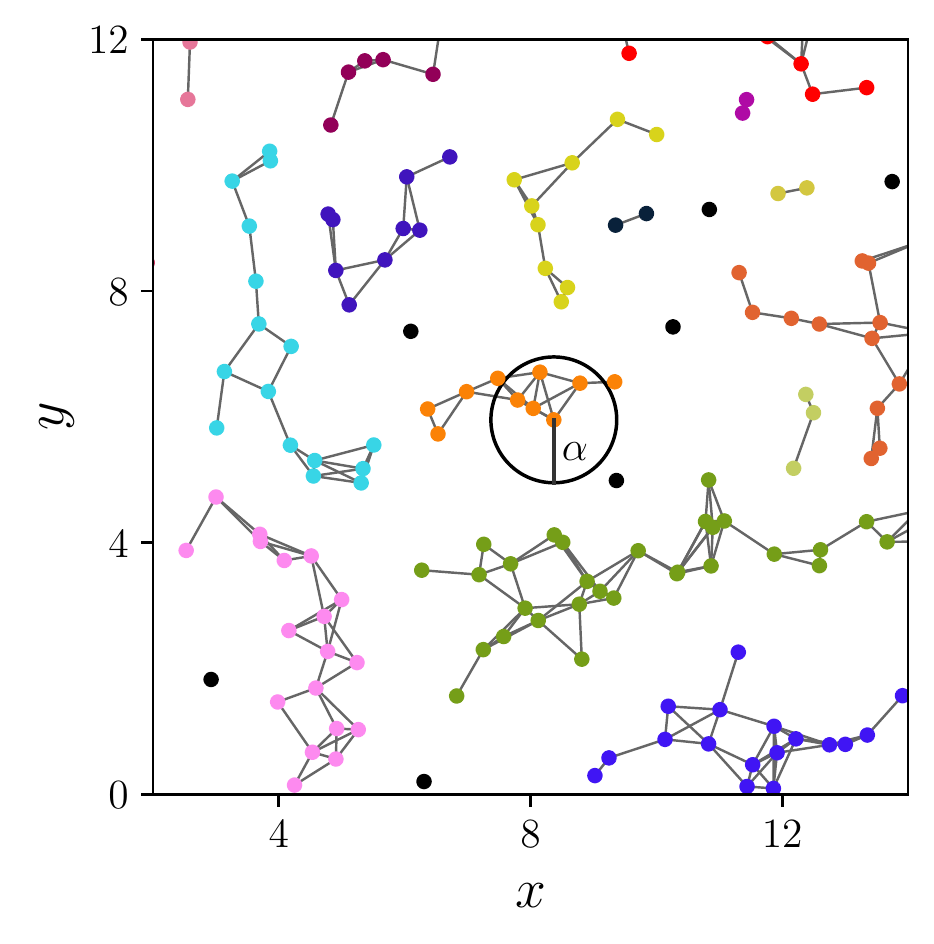}
\caption{Example of a RGG with $M=256$ nodes, $d=1.0$ and $r=\alpha = 1.0$. The square box has a linear size $L = \sqrt{M} = 16$. Only a part of the area is represented in the figure in order to emphasize the neighborhood radius. A complete square is shown on the Fig. \ref{RGG_comps}. Each color represents a different component and the isolated nodes (without neighbors) are in black. The circle with radius $\alpha$ defines the neighborhood of influence of the node in its center.}
\label{RGG}
\end{figure}

The criterion to define the connectivity is the following: two sites are connected, i.e., they are neighbors, if and only if the euclidean distance between them is less than $\alpha$. An isolated set of nodes connected to each other defines a component \cite{Newman2010} as represented in Fig. \ref{RGG} (each color is a different component). In other words, a component of a graph is a group of nodes such that it is possible to go from any node to any other node through the connections among them. A graph with only one component is called connected, which means there is no isolated nodes. Therefore, the spatial positions of the nodes are the criteria to define the network connectivity of the RGG.

The number $k_i$ of neighbors of a given site $i$ is called the degree and the average network degree is given by
\begin{equation}
\bar{k}=\sum_i^M \frac{k_i}{M}. \nonumber
\end{equation}
The value of $\bar{k}$ for the RGG can vary from 0, when $\alpha = 0$ and the system is totally disconnected, to $ M-1$ when $\alpha > \sqrt{2}L/2$ (half the diagonal of the square) and the system is fully connected, i.e., everyone is everyone's neighbor. This particularity brings an important consequence: for small values of $\alpha$, the network is not connected and there exist many components. The number of components $\mathcal{N}$ depends on the radius $\alpha$ since when it increases, $\bar{k}$ also increases and, consequently, the components grow up in size decreasing the value of $\mathcal{N}$. So, $\mathcal{N}$ can vary from $M$ when $\bar{k} = 0$ meaning a totally disconnected system, to 1 when $\bar{k}=M-1$ and the network is fully connected. 

Another important quantity to be considered is the size of the largest component $\mathcal{S}$, i.e., the number of sites belonging to the largest component of the network. This quantity can vary from 1 meaning that all nodes are isolated ($\bar{k}=0$) to $M$ when the network is fully connected (only one component, which is the size of the network) \cite{Reia2020}. Therefore, for a network with a certain set of components, the size of its largest component can bring relevant information about the connectivity of the nodes. For instance, there exists a radius threshold in which the largest component $\mathcal{S}$ percolates and the network gets connected \cite{dall2002}. In order to study the phase transitions of the ZGB model on this random network, the radius threshold becomes an important parameter from which the phase transitions of the system can occur.

\subsubsection{The Erd\"{o}s R\'{e}nyi network (ERN)}

The Erd\"{o}s R\'{e}nyi network (ERN) is also constructed by placing $M$ nodes in the network at random. However, the connections are chosen according to the probability $p$, i.e., each node has a probability $p$ to be connected to another one \cite{Newman2010,Barabasi2016}. Hence, there is no spatial position in the definition of the ERN. This is the main difference between the RGG and the ERN: for a given number $M$ of nodes, while the neighborhood radius $\alpha$ is the control parameter for the RGG, the probability $p$ is the control parameter for the ERN. The other quantities such as $\mathcal{N}$, $\mathcal{S}$ and $\bar{k}$ have the same meaning for the two networks. Therefore, the probability $p$ is responsible for producing a network ranging from totally disconnected ($p = 0$ and $\bar{k} = 0$) to fully connected ($p=1$ and $\bar{k} = M-1$). In the general case we have $\bar{k} = p(M-1)$. As it was shown above for the RGG, there is also a critical value of $p$ and $\bar{k}$ that makes the largest component large enough to allow the percolation of the system \cite{Bolobas2001,erdos1959}: $p_c = 1/M$ which implies $\bar{k}_c=1$. If $p>p_c$ the network is percolated \cite{Achlioptas2009}. Indeed, it is needed at least one neighbor for each site in order to have a percolated graph.

For regular SL, the edges are the nearest-neighbors pairs of sites so that $k_i=4$ for all site $i$. Hence, the lattice is always connected (only one component): $\mathcal{N}=1$, $\mathcal{S}=M$.

\section{Results and discussion} \label{results}

\subsection{Monte Carlo simulations}

Before presenting our main results, we make a few general considerations about the Monte Carlo simulations carried out in this work, as well as, present the main topological properties of the RGG and ERN.

The model was simulated on catalytic surfaces with $M=128^2$ sites with periodic boundary conditions. We used the density $d=1.0$ throughout this paper, which gives $\ell=1.0$ and $r=\alpha$. During the simulations, we compute the densities of CO molecules, $\rho_c$, and O atoms, $\rho_o$, both adsorbed on the catalytic surfaces, as well as the density of vacant sites, $\rho_v$, and the density of CO$_2$ molecules, $\rho_2$, released from the surface during the catalytic reaction. These densities are given by:
\begin{eqnarray}
\rho_c &=& \dfrac{n_c}{M}, \qquad \qquad \rho_o = \dfrac{n_o}{M}, \nonumber \\
\rho_2 &=& \dfrac{n_2}{M}, \qquad \qquad \rho_v = \dfrac{n_v}{M}, \nonumber
\end{eqnarray}
where $n_c$, $n_o$, $n_2$ and $n_v$ are, respectively, the number of sites occupied by CO molecules, O atoms, the number of CO$_2$ molecules produced in the active phase and number of empty sites.

The basic steps of the simulation algorithm are:
\begin{enumerate}
\item Definition the input parameters $M$, $y$, $\alpha$ or $p$, $A$ and $Q$.
\item Creation one of the networks (RGG or ERN) and calculation of its topological properties ($n$,$s$,$\bar{k}$) as function of $\alpha$ or $p$, respectively.
\item Selection of a CO or O$_2$ molecule in the gas phase with probability $y$ and $1-y$, respectively, as well as one site on the catalytic surface at random. Perform the adsorption process and the catalytic reaction of the model as shown in Eqs. \ref{reaction1}, \ref{reaction2} and \ref{eq:reaction}. 
\item Evaluation of the steps above $M$ times is one Monte Carlo step (MCS). 
\item Thermalization is achieved after $Q$ Monte Carlo steps and the system is considered to be in steady state.
\item Calculation the averages of interest ($\rho_v$,$\rho_c$,$\rho_o$,$\rho_2$) over the next $A$ Monte Carlo steps.
\end{enumerate}

The simulations started with all sites empty (initial condition): $\rho_c = \rho_o = 0.0$ and $\rho_v = 1.0$ (which implies $\rho_2=0.0$). We discarded the first $Q=8.0 \times 10^3$ Monte Carlo steps (termalization process) in order to reach the steady state. The number of samples, where the results (densities) are averaged, is $A=1.5\times 10^4$ MCS for the RGG and $A=2\times 10^4$ MCS for the ERN. All the simulations were carried out using Fortran and all the graphics were created using Python.

\subsection{Topological properties of the networks}

To study the two random networks, we considered three topological properties: the average degree distribution $\bar{k}$, as well as the normalized number of components $n$ and largest component $s$ given by
\begin{equation}
    n = \dfrac{\mathcal{N}}{M}, \qquad \qquad \textrm{and} \qquad \qquad s = \dfrac{\mathcal{S}}{M}. \nonumber
\end{equation}
So, for a totally disconnected network we have $\mathcal{N}=M$ and $\mathcal{S}=1$ producing $n=1$ and $s=1/M$. On the other hand, for a fully connected network $\mathcal{N}=1$ and $\mathcal{S}=M$ which gives $n=1/M$ and $s=1$.

\begin{figure}[h!]
\centering
\includegraphics[width=3.05 in]{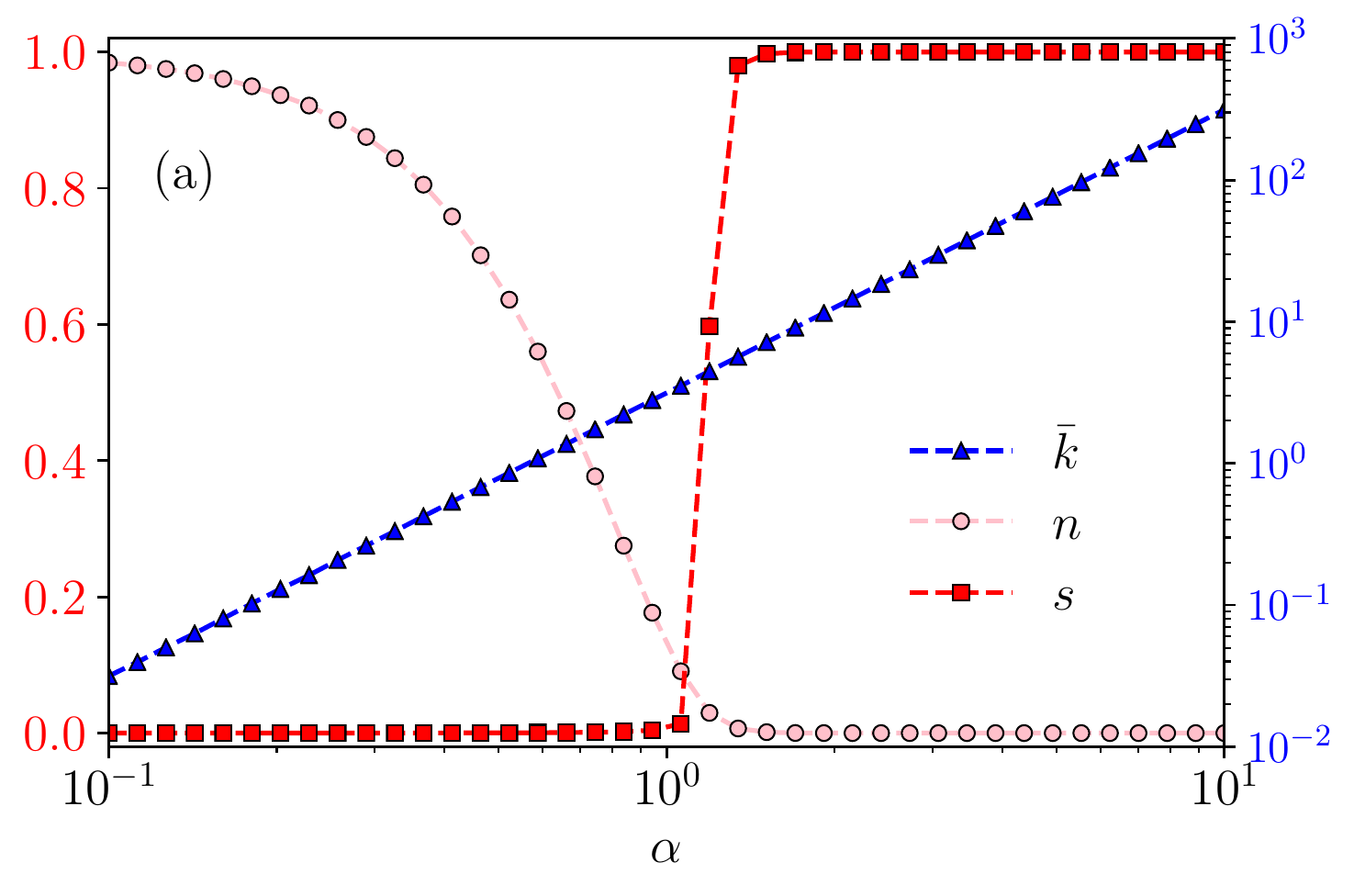}
\includegraphics[width=3.0 in]{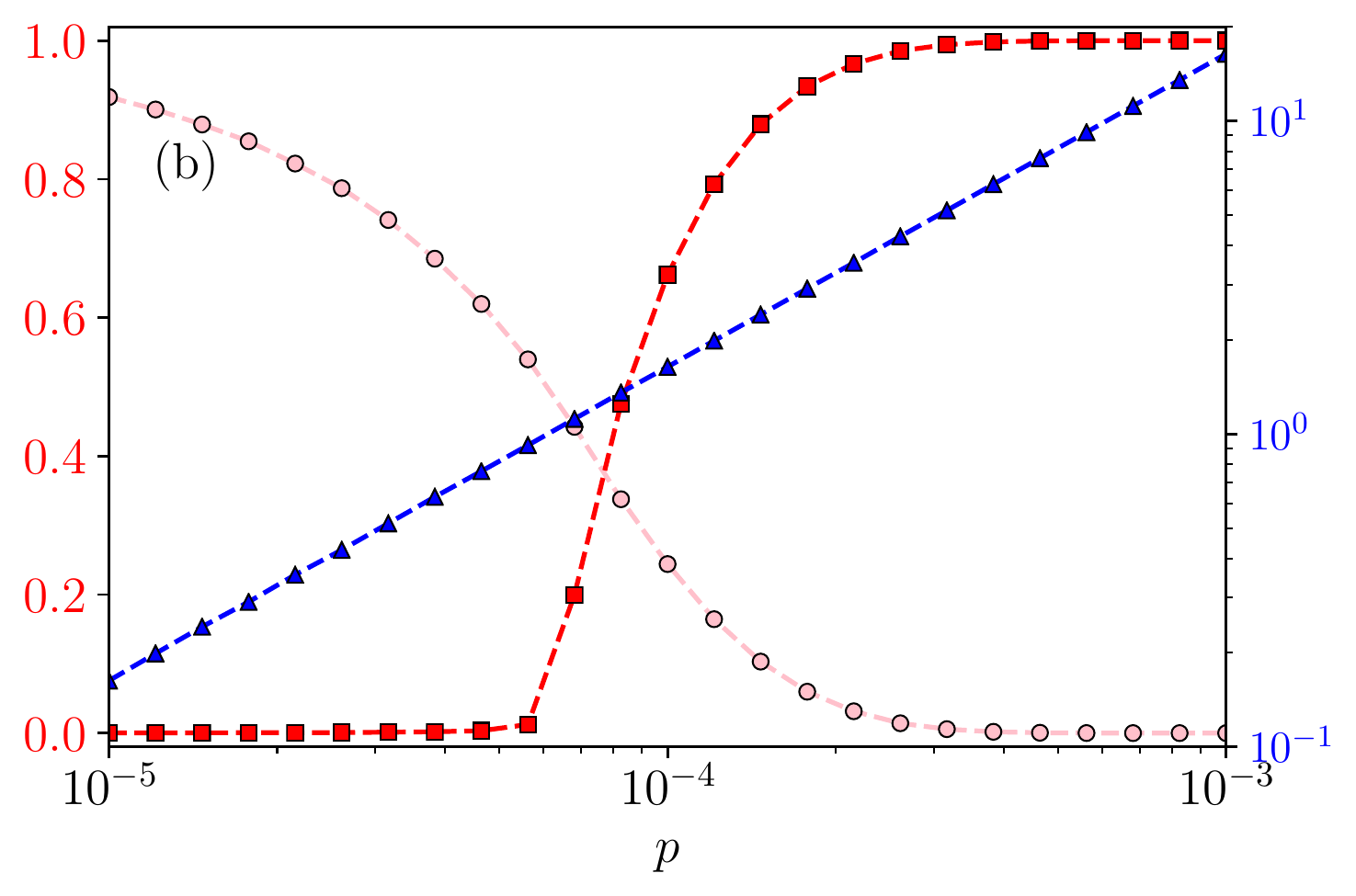}
\caption{Topological properties as function of the control parameters for (a) RGG and (b) ERN. The curves show the average degree distribution $\bar{k}$ (in blue triangles), the number of components $n$ (in pink circles), and the largest component $s$ (in red squares). The dashed lines are just a guide to the eye.} 
\label{topological}
\end{figure}

Figures \ref{topological}(a) and (b) show these three quantities for the RGG and ERN, respectively. These results were obtained in the interval $0.1 \leq \alpha \leq 10$ and $10^{-5} \leq p \leq 10^{-3}$ which are centered on the percolation point. As can be seen in this figure, $\bar{k}$ has linear behavior in log scale for both networks.
Another consequence of the degree distributions is that the random networks do not percolate for all values of the control parameters $\alpha$ and $p$. The percolation begins when the system is close to be connected meaning that there are only a few components.

As shown in Fig. \ref{topological}(a), the percolation starts for $\alpha \gtrsim 1.0$ since $s$ goes from $\simeq 0$ to $\simeq 1.0$ suddenly around this point meaning that even a small variation of $\alpha$ can produce huge transformations on the network from disconnected to connected. The normalized number of components $n$ can also provide important information about the graph percolation even though it decreases smoothly from one to zero and. At $\alpha \simeq 1.0$ its value is $n \sim  0.1$ meaning that only 10\% of the nodes are not connected to the largest component.

A similar behavior is found for the ERN (see Fig. \ref{topological} (b)) although the curve of $s$ is not as abrupt as for the RGG. We found that the percolation process is centered around $p \simeq 10^{-4}$ when 70\% of the nodes ($n=0.3$) are connected to the largest component. It starts at $p  \sim 6 \times 10^{-5}$, which is in accordance with the predicted value $p_c = 1/M \sim 6.1 \times 10^{-5}$ \cite{Barabasi2016}. The average degree at this point is also in agreement with the prediction $\bar{k}_c=1$. These topological features will be important for the description of the phase diagrams, shown in the following sections.

\subsection{Densities and phase transitions}

\begin{figure*}
\centering
\includegraphics[width=2.3 in]{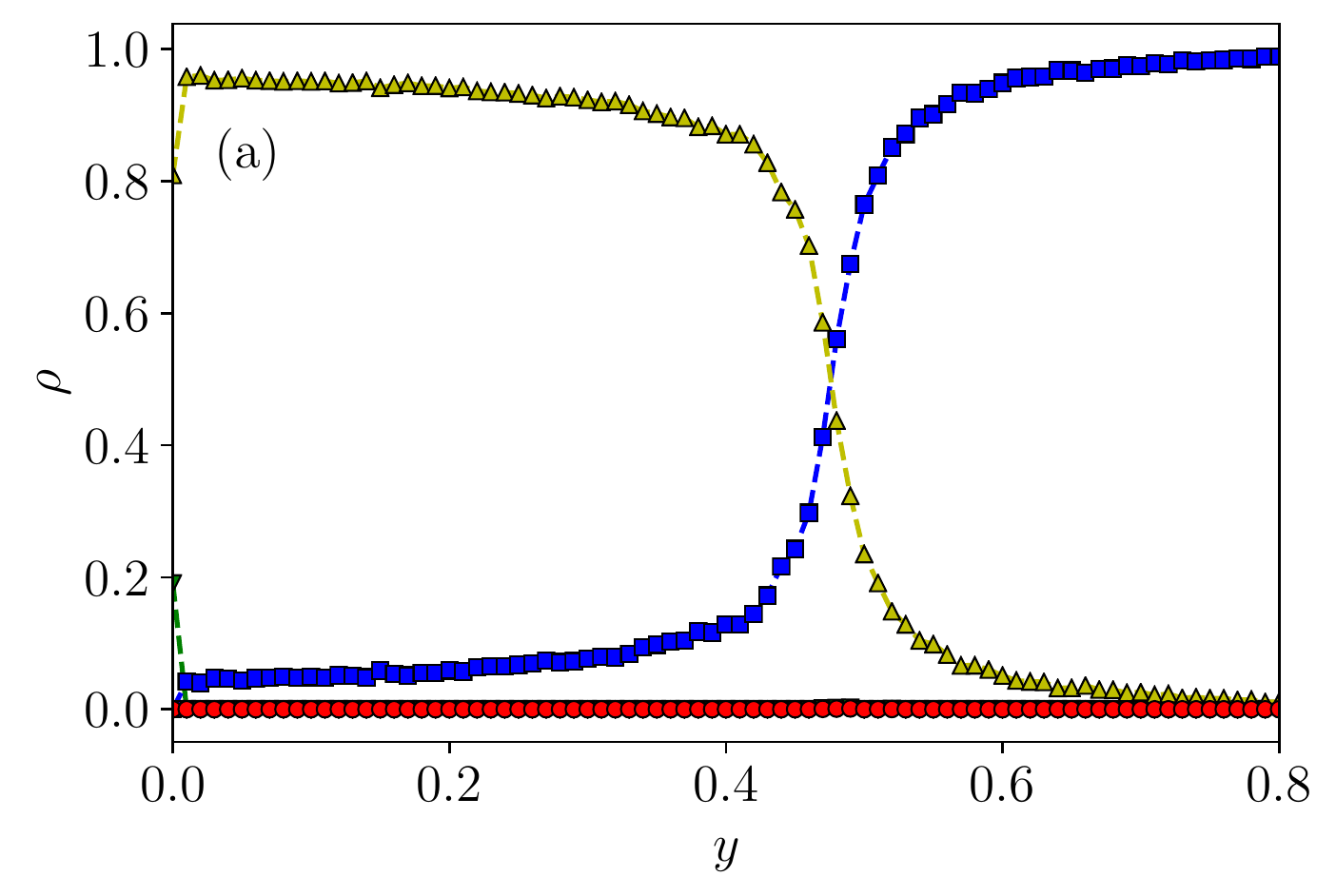}
\includegraphics[width=2.3 in]{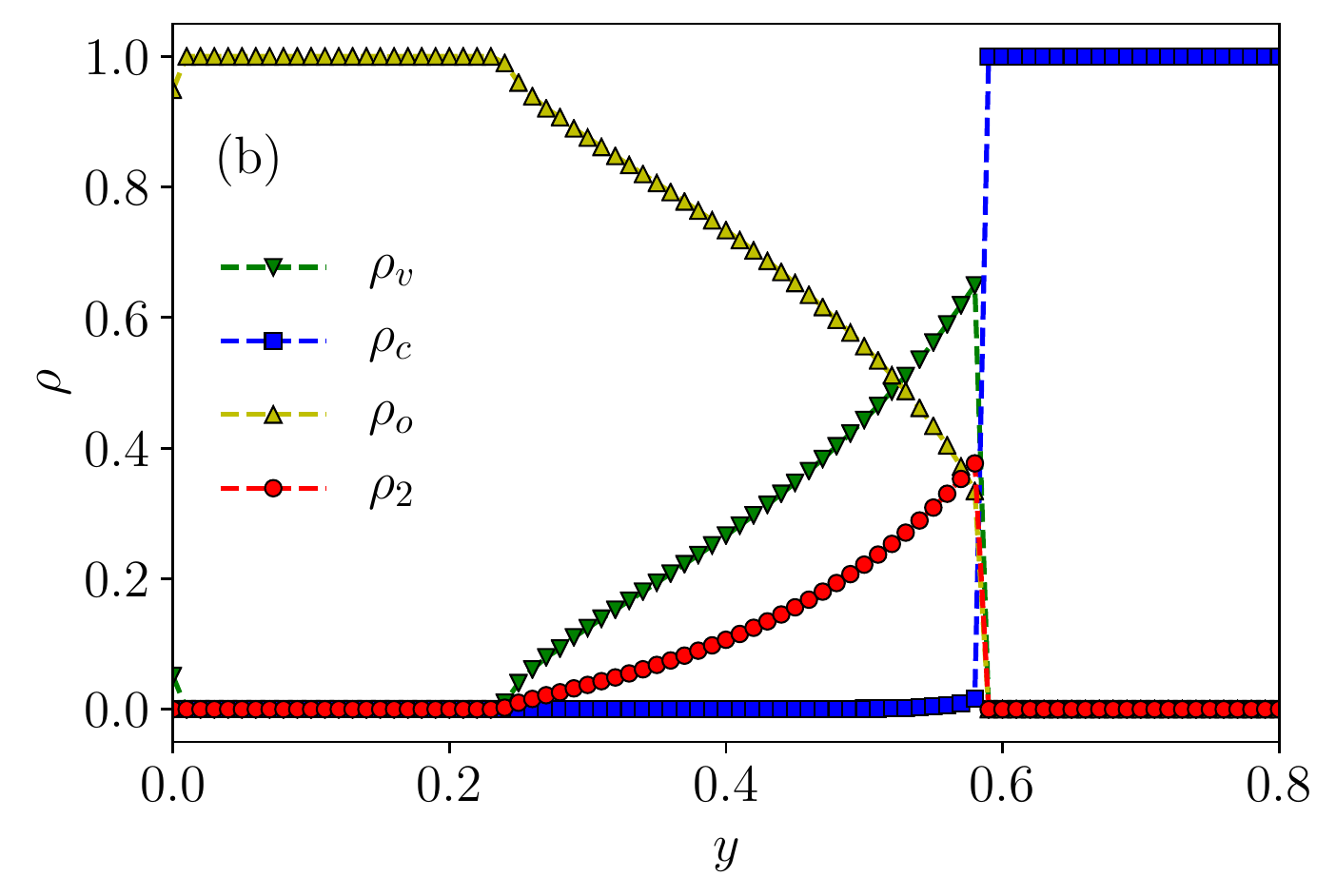}
\includegraphics[width=2.3 in]{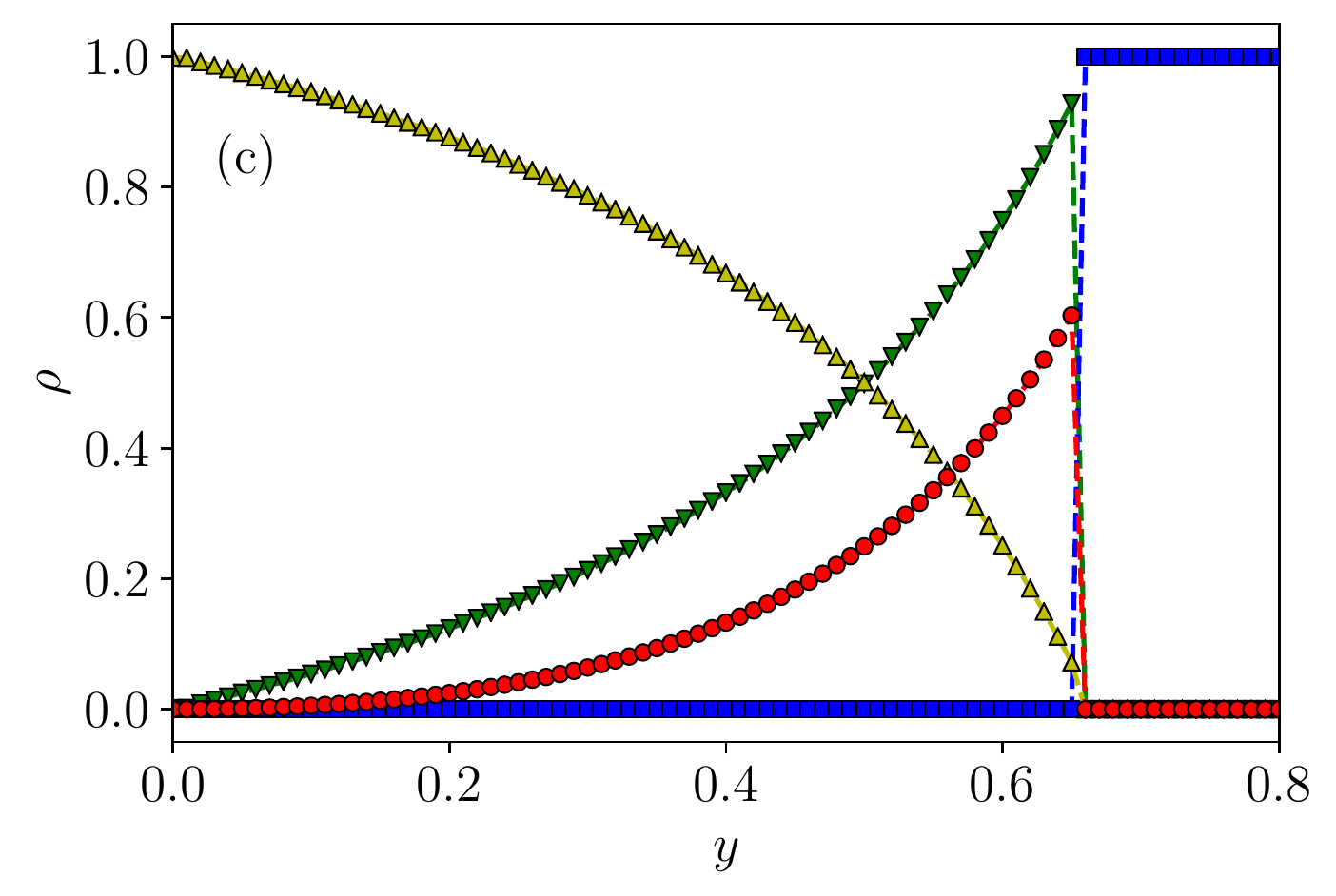}
\caption{Phase diagrams of the ZGB model on the RGG with $M=128^2$, $d=1.0$ and for different values of $\alpha$. (a) $\alpha=1.0$, $\bar{k} \approx 3.13$, $n \approx 0.131$ and $s \approx 0.013$. (b) $\alpha=2.0$, $\bar{k} \approx 12.6$, $n \approx 0.0$ and $s \approx 1.0$. (c) $\alpha=12.0$, $\bar{k} \approx 452.4$, $n \approx 0.0$ and $s \approx 1.0$. The densities $\rho_v$, $\rho_c$, $\rho_o$, and $\rho_2$ are represented by green up side down triangle, blue square, yellow triangle, and red circle, respectively. The dashed lines are just a guide to the eye. The analogous results for the ERN are available on the supporting information.} 
\label{densidadesRGG}
\end{figure*}

At first, we compare the behavior of the densities as function of $y$ for the ZGB model on the RGG and ERN with those on the SL. The curves for the two random networks are very similar and, therefore, we decided to present only the results for RGG. In Fig. \ref{densidadesRGG}, we present our findings for three different values of $\alpha$.


As shown in Fig. \ref{densidadesRGG}(a), for low values of $\alpha$ ($\alpha = 1.0$), the system is in a regime without production of CO$_2$ molecules for any value of $y>0$. In this case, the catalytic surface is composed by more than two thousand components, each one with a unique particle species: or O atoms or CO molecules. For $y=0$, there exist only O atoms (when the network has components with two or more nodes) and vacant sites. For low values of $y$, most components are filled with O atoms but some isolated nodes or even larger components may have CO molecules. As $y$ increases, the number of components with CO molecules also increases, as it should, and for $y \simeq 0.5$ the number of CO molecules grows rapidly while $\rho_o$ decreases. This phase diagram shows that the network is in an absorbing state for all values of $y$. However, this state is not as the two absorbing ones of the original ZGB model since there are no O- or CO-poisoned states. Instead, this absorbing phase is composed from O atoms, CO molecules and vacant sites. For this reason, we name this phase state as inactive state. 

Figure \ref{densidadesRGG}(b) shows that the framework is completely different for $\alpha=2.0$ since the phase diagram now resembles the original one of the ZGB model: two poisoned states separated by an active phase with production of CO$_2$ molecules. However, the window of the active phase is larger than that of the standard model. The point separating the O-poisoned phase from the active state is $y \lesssim 0.24$ and the point separating the active phase from the CO-poisoned phase is $y \gtrsim 0.58$. It is worth to mention that, for this value of $\alpha$, the number of components $n$ decreases substantially (see Fig. \ref{topological}(a)) approaching zero meaning that the network is already connected and percolated. That figure also shows that for this value of $\alpha$ the average degree is $\bar{k}\simeq 12.6$ and $s\simeq 1$. 

The phase diagram remains with three well-defined phases and the reactive window grows up until $ \alpha \lesssim 12$. From this point on, the O-poisoned state disappears for $y>0$ and the only present phase transition  is the discontinuous one. Figure \ref{densidadesRGG}(c) shows that for $\alpha=12.0$ there is no O-poisoned state for any value of $y$ greater than zero preventing the model to have the continuous phase transition. On the other hand, the discontinuous phase transitions occurs at $y \simeq 0.65$. For this value of $\alpha$, $\bar{k} \simeq 452$, such that the most $O_2$ molecules which impinge the surface are adsorbed on it. In addition, the catalytic reaction, i.e., the CO$_2$ production, is greatly increased even for low values of $y$ by the amount of neighbors of each O atom adsorbed after the dissociation process.

The behaviour of these densities in the ERN (not shown) are completely analogous to the ones described above for RGG. The inactive state in the ERN can be observed for $p \lesssim 10^{-4}$ although a small production of CO$_2$ starts when the system percolates at $p =6.1 \times 10^{-5}$. The two poisoned states and the active phase can be observed at $p=5\times 10^{-4}$ . In this configuration the active phase range is $0.19 \lesssim y \lesssim 0.61$. The last configuration where the O poisoned states is absent occurs at $p \gtrsim 10^{-2}$, such that at $p = 10^{-1}$ we obtain an almost identical behaviour as the one presented at Fig. \ref{densidadesRGG}(c). 
 
\begin{figure*}
\centering
\includegraphics[width=3.1 in]{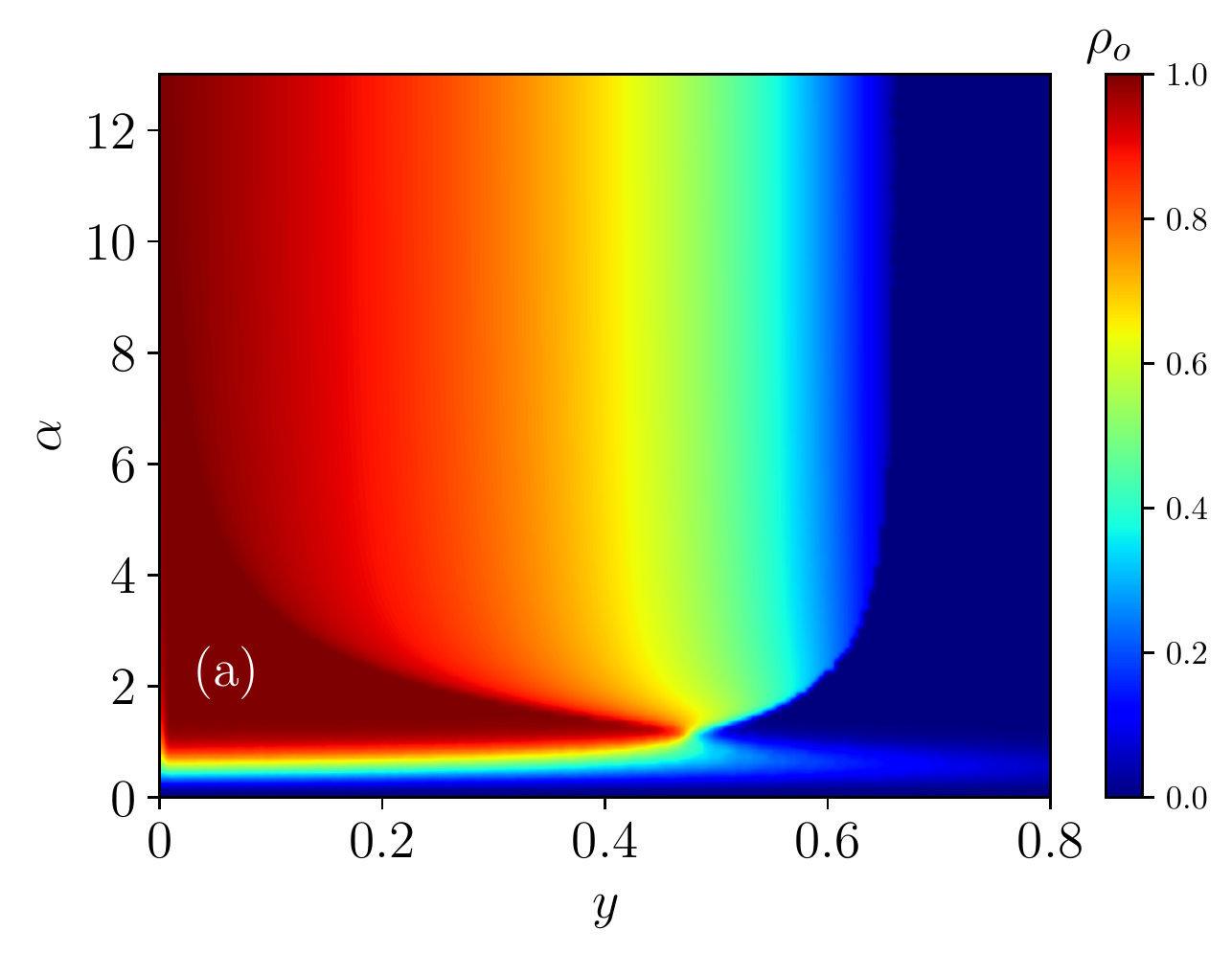}
\includegraphics[width=3.1 in]{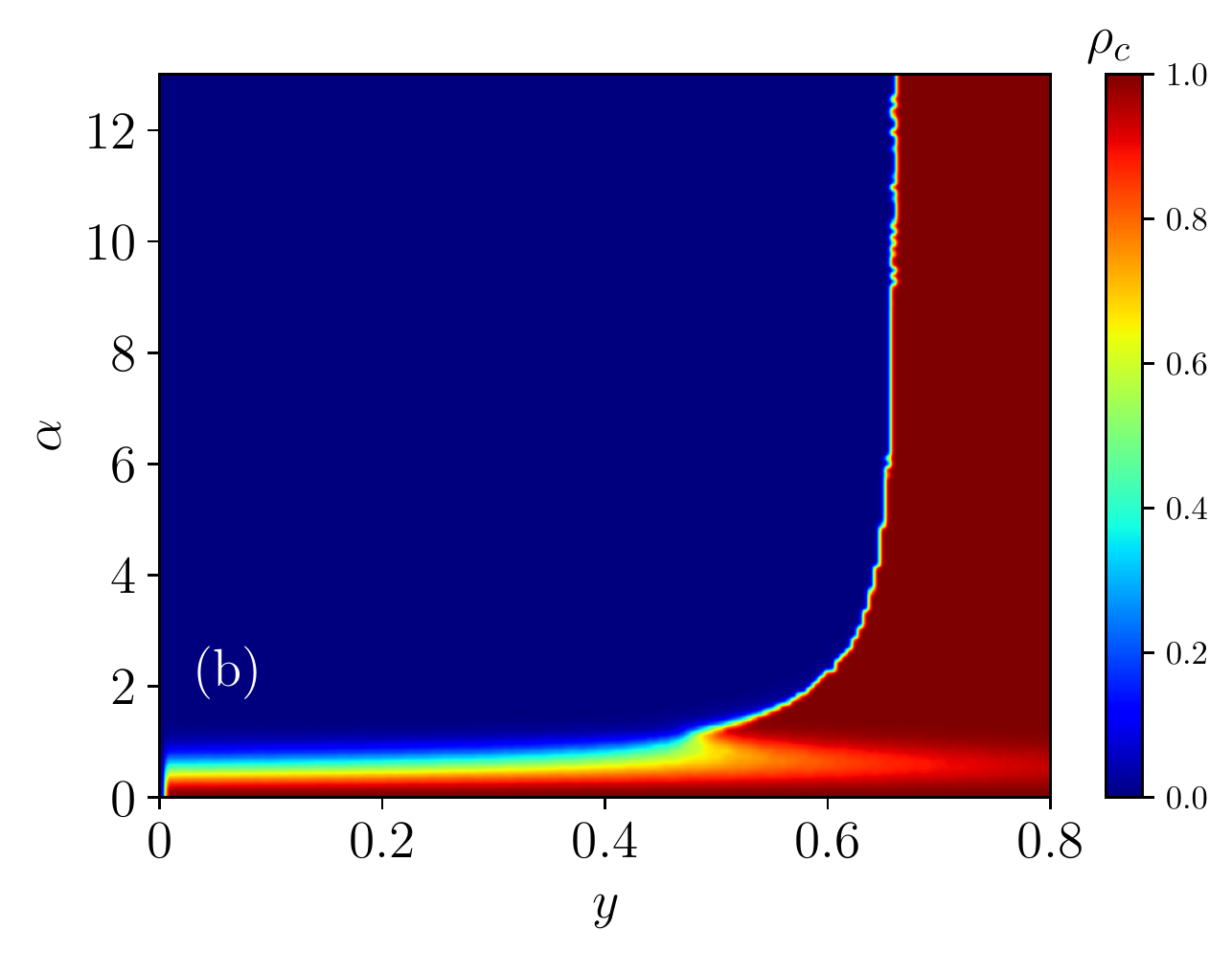} \\
\includegraphics[width=3.1 in]{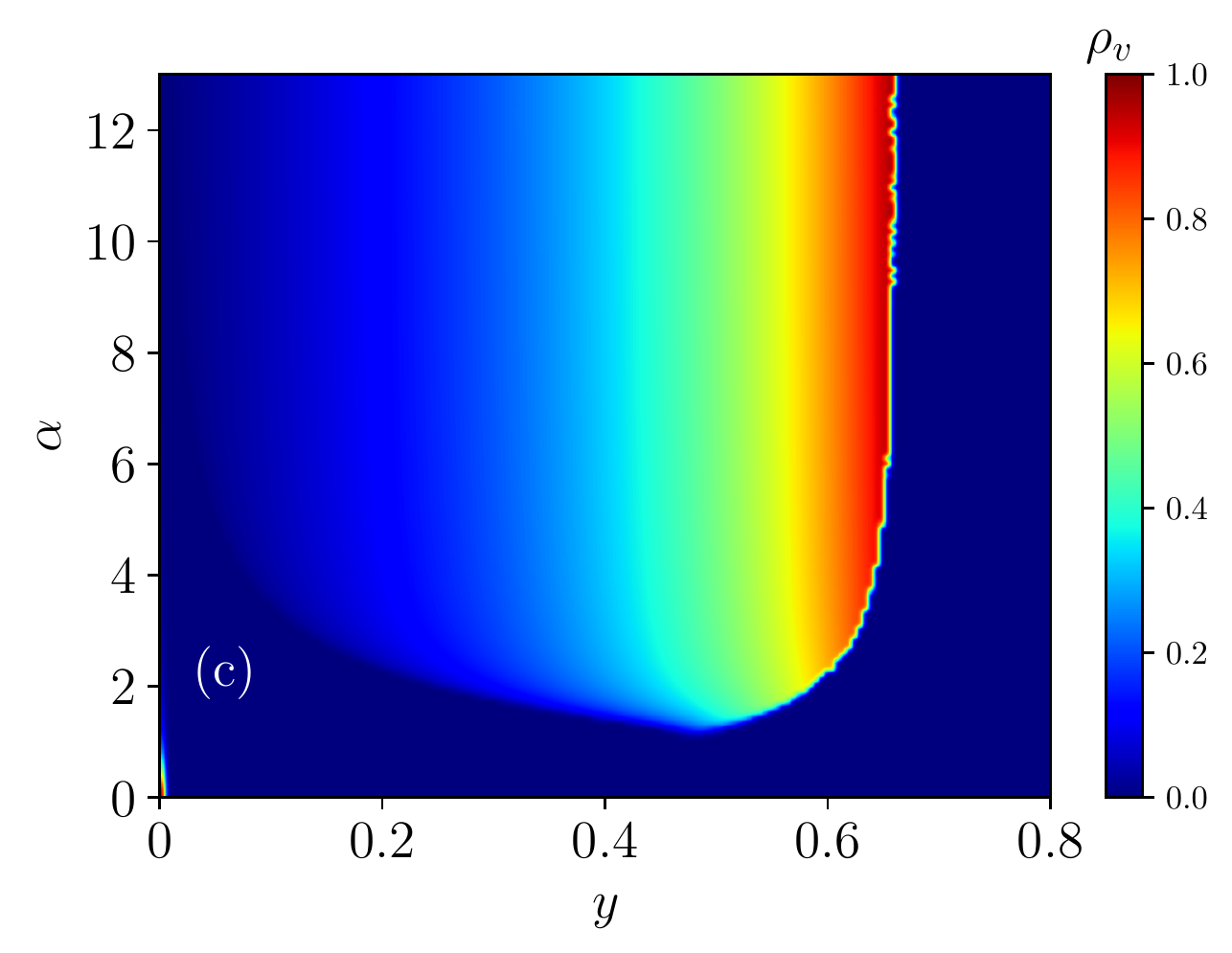}
\includegraphics[width=3.1 in]{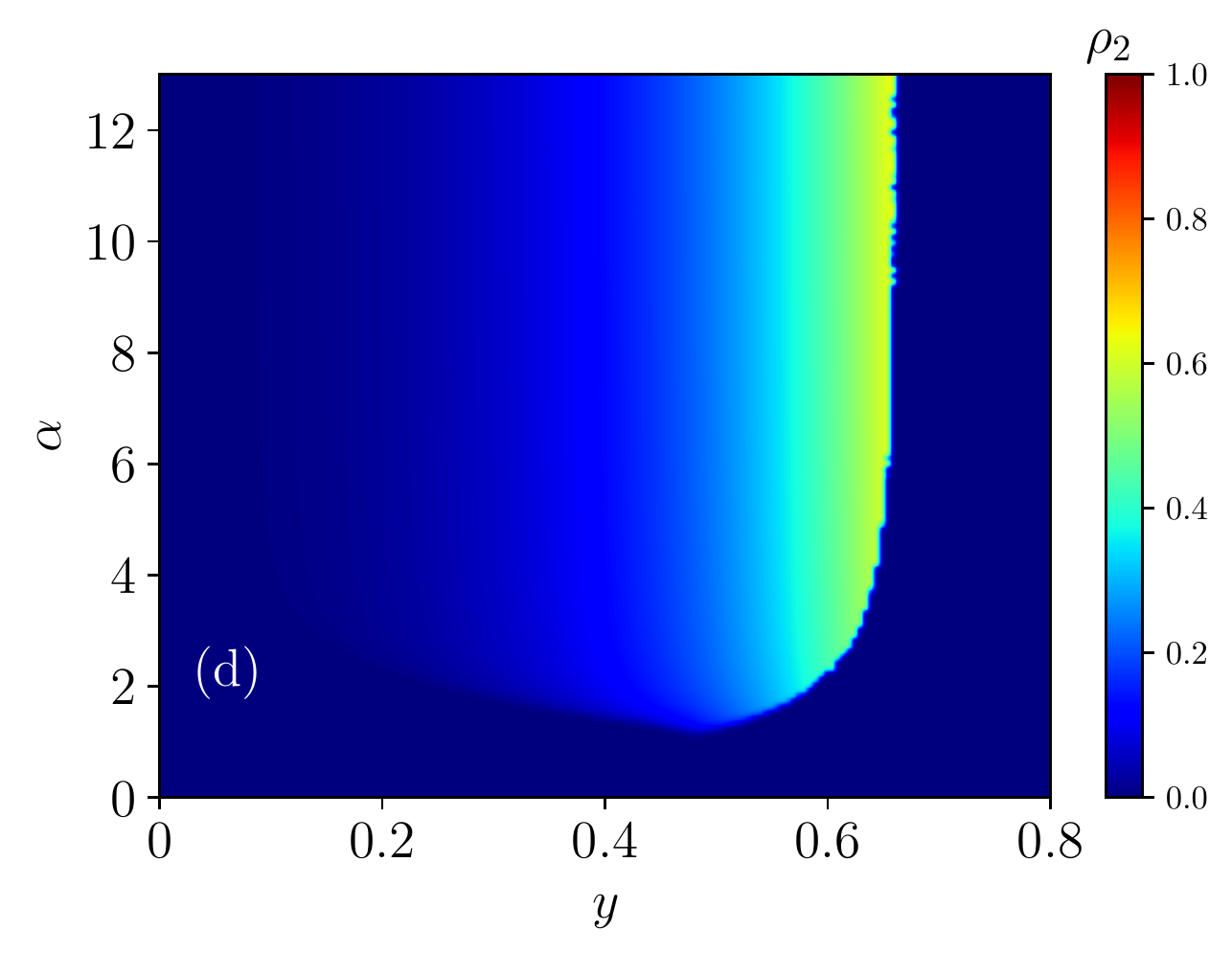}
\caption{Color map of the densities (a) $\rho_o$, (b) $\rho_c$, (c) $\rho_v$, and (d) $\rho_2$ as function of $\alpha$ and $y$ for the RGG. The steps are $\Delta y=0.005$ and $\Delta \alpha=0.1$. Parameters: $M=128^2$ and $d=1.0$. The analogous results for the ERN are available on the supporting information.}
\label{mapa_rho}
\end{figure*}

For each random network there is a control parameter ($\alpha$ for RGG and $p$ for ERN) which defines its average degree $\bar{k}$, the distribution of components ($n$ and $s$) and all other topological features. On the model side, the only control parameter is the CO adsorption rate $y$, therefore, only two parameters define the state of the system in each case. For this reason, we are able to construct color maps showing the densities as function of ($\alpha,y$) for the RGG and ($p,y$) for the ERN. However, we also suppress the results for the ERN since they are very similar to the RGG. Figure \ref{mapa_rho} presents the density color maps for the RGG in the range $0<y<0.8$ and $0<\alpha<12$. 

In Fig. \ref{mapa_rho}(a), we present the color map of $\rho_o$ in order to show the regions of the parameter space in which there exist O atoms. As can be seen the O-poisoned phase, which occurs whenever $\rho_o=1.0$, is indicated by the dark red color at the left side of the figure, which holds on only for low values of $\alpha$ and $y \lesssim 0.5$. For $\alpha \longrightarrow 0$ the only possible molecule to be adsorbed on the surface is the CO molecule (blue color) since the network has only isolated components. On the other hand, as stated above, for large values of $\alpha$, the chance of an O atom adsorbed on the surface to react with one CO molecule is high enough to prevent the poisoning of the network and so, $\rho_o < 1.0$. This figure also shows that for $y \gtrsim 0.65$ the density of O atoms drops to zero for any value of $\alpha$ and, as presented in Fig. \ref{mapa_rho}(b), this is the region where the system reaches the CO-poisoned state defined by $\rho_c=1.0$. This state, indicated by the dark red color, can be observed for $y \simeq 0.5$ and low values of $\alpha$ as well. Figure \ref{mapa_rho}(c) presents the density of vacant sites $\rho_v$. In this plot we can see that the dark blue color is related to the poisoned states in which there is no empty sites and $\rho_v=0$. On the other hand, one can notice a red colored region for $\alpha \gtrsim 0.3$ and $y$ around 0.6, which in turn indicates that each site has a large number of neighbors as well as there are more CO molecules impinging the surface than O$_2$ molecules. So, almost every O$_2$ molecule which reaches the surface is absorbed on it, interacting with CO molecules and creating two CO$_2$ molecules. Finally, in Fig. \ref{mapa_rho}(d) one can see that the region of greatest production of CO$_2$ molecules is also the region of higher number of vacant sites. We can also observe the system approaching a stable regime at $\alpha \gtrsim 10$ in which there is no more continuous phase transition and the point $y$ of the discontinuous one becomes constant.

\subsection{CO$_2$ production}

\begin{figure}
\centering
\includegraphics[width=3.0 in]{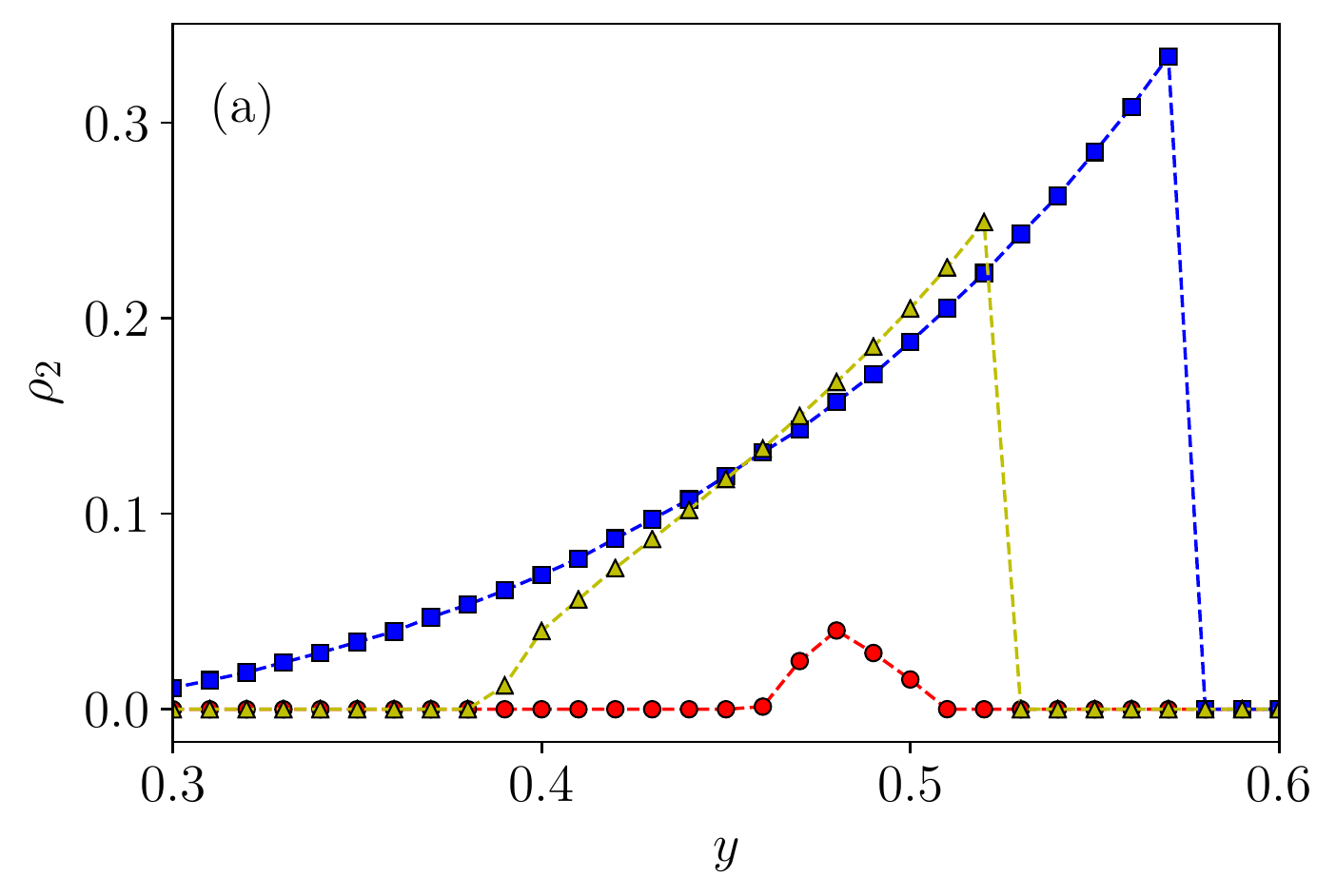}
\includegraphics[width=3.0 in]{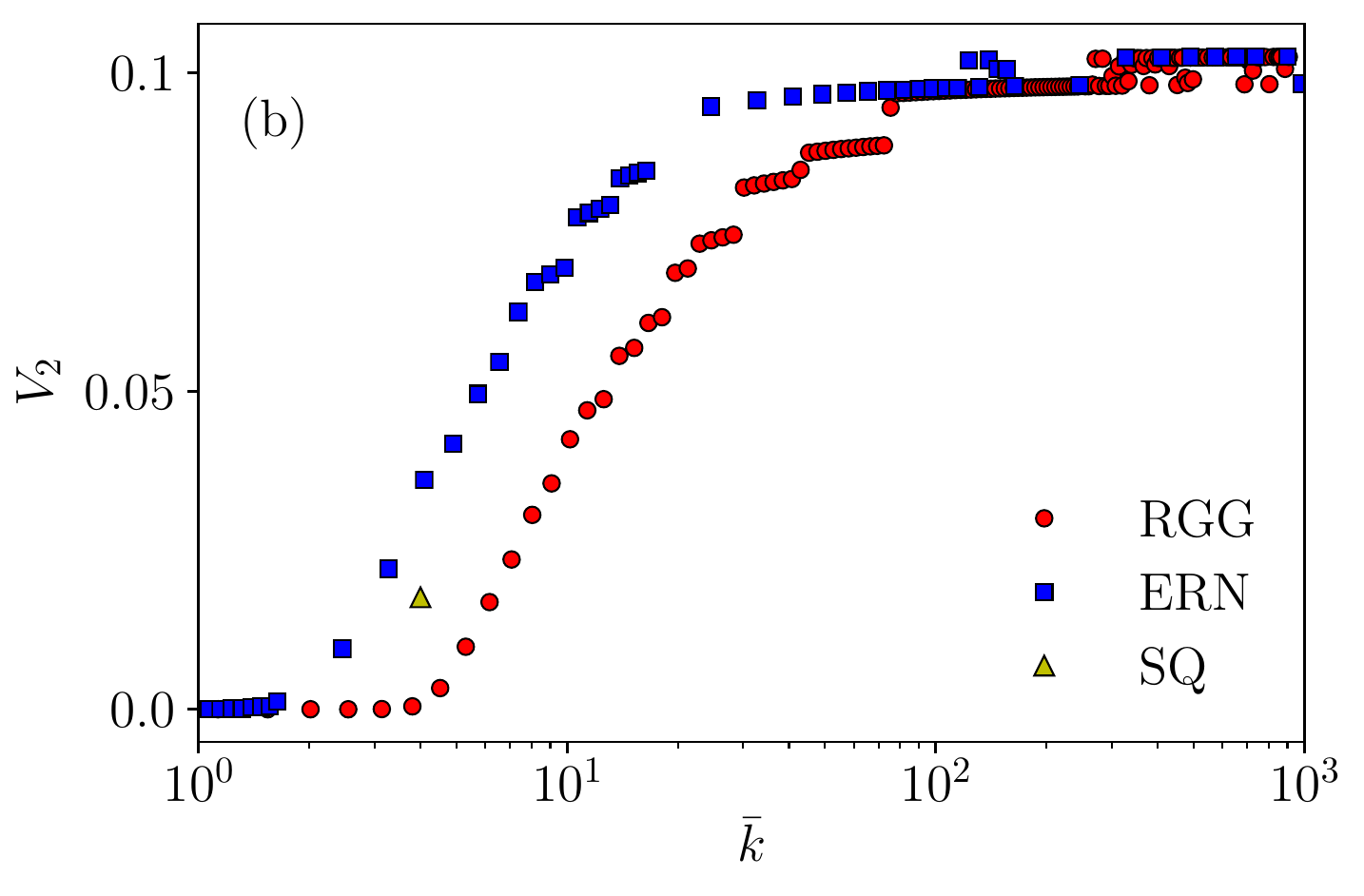}
\caption{(a) Density $\rho_2$ vs $y$ with $M=128^2$ and $d=1.0$ for the two random networks with $k \sim 4$ compared to the standard square lattice. The dashed line is just a guide to the eye. (b) Volume of total production of CO$_2$ $V_2=\int \rho_2 dy$ as function of $\bar{k}$ for the two random networks and for the square lattice. RGG: red circle. ERN:  blue square. SL: yellow triangle.}
\label{comparacao_RN_SQ}
\end{figure}

In order to compare the production of CO$_2$ molecules between the two random networks considered in this study with the traditional SL, we chose the parameters $\alpha$ and $p$ so that $\bar{k} \simeq 4$, thus creating networks whose sites have a comparable number of neighbors to those of the SL ($k=4$). Figure \ref{comparacao_RN_SQ}(a) shows the CO$_2$ density $\rho_2$ as function of $y$ for the three networks.
Although the connectivity of both random networks is almost the same, Fig. \ref{comparacao_RN_SQ}(a) shows a striking discrepancy in the behavior of $\rho_2$. While ERN produces a lot of CO$_2$ molecules for a large window of $y-$values, the active phase of RGG is restricted to a narrow window and with little production of CO$_2$. In addition, the production of CO$_2$ molecules on the ERN resembles the one on the regular SL even though the continuous phase transition is not observed on the ERN with $\bar{k}\simeq 4$. This result shows that although the connectivity is the first parameter considered when using different networks it is not the only important one to determine their properties. The effect of the topological properties of the networks can be more explicitly seen in the integrated intensity of $\rho_2$ as function of the average degree $\bar{k}$, which is proportional to the total volume of CO$_2$ produced in the range $0<y<1.0$. Figure \ref{comparacao_RN_SQ}(b) displays the greater efficiency in CO$_2$ production for the ERN when compared to the RGG at $\bar{k} < 100$ while at $\bar{k}>100$ both networks have similar productions. The SL is represented by the yellow triangle at $\bar{k}=4$. At this point, ERN is twice as more efficient than SL and the production of CO$_2$ through RGG is close to zero. This is a dramatic evidence of the importance of the network topological properties on the system.

\subsection{Phase diagrams}

\begin{figure}
\centering
\includegraphics[width=3.1 in]{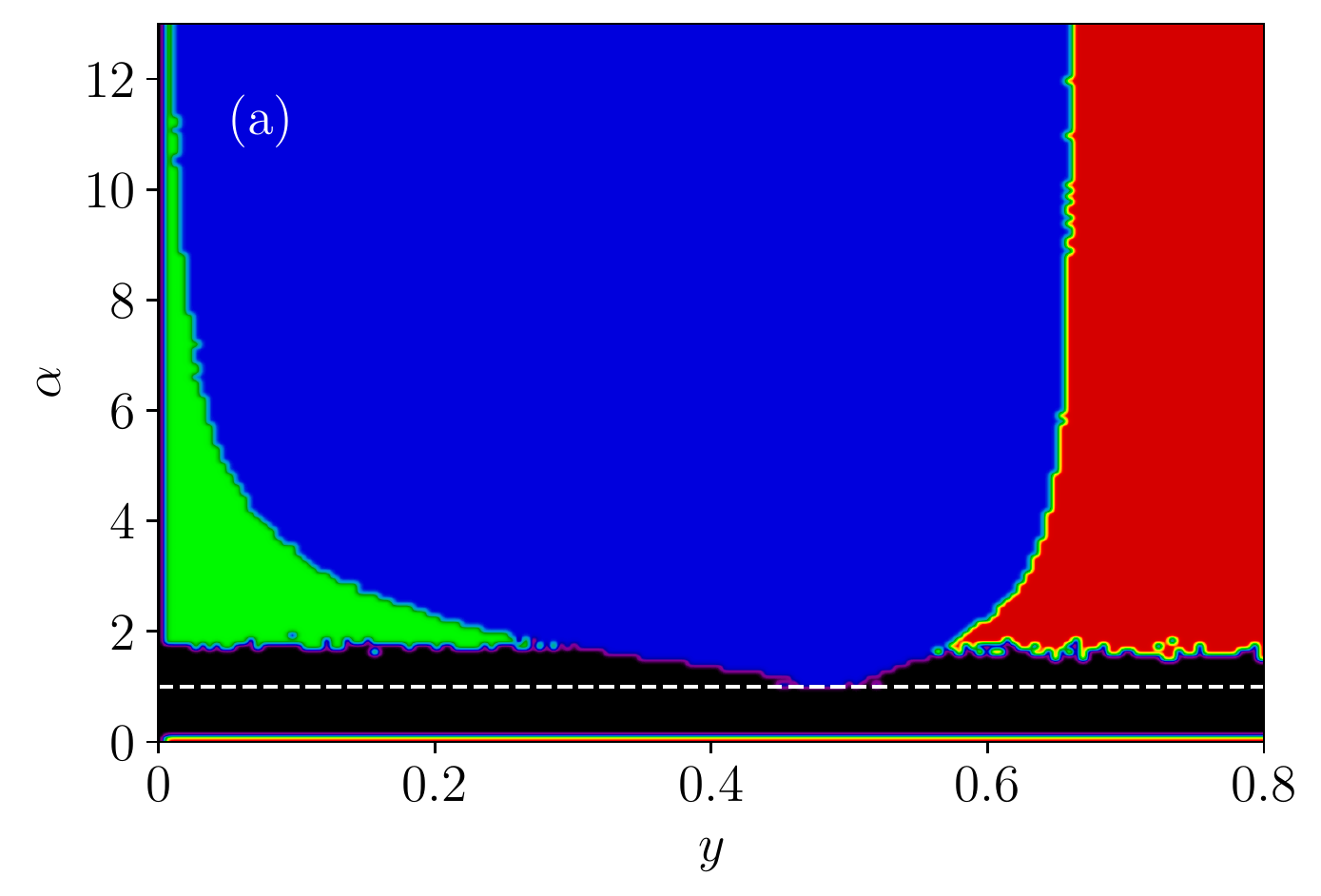}
\includegraphics[width=3.2 in]{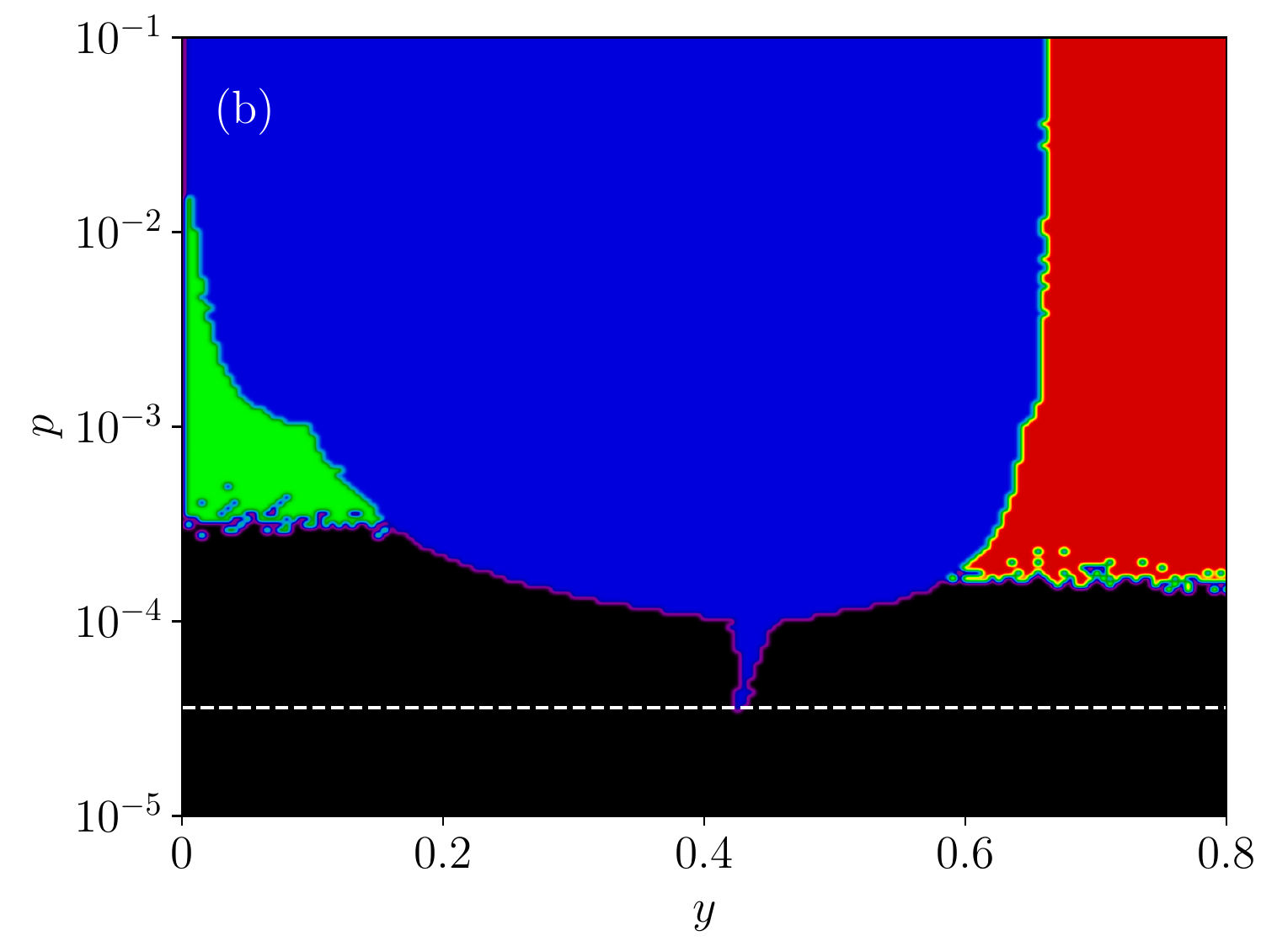}
\caption{Phase diagrams of the ZGB model on the (a) RGG and (b) ERN. Blue region represents the active phase ($\rho_2>0$) where CO$_2$ is produced. The red region is the CO-poisoned phase ($\rho_c>0.9999$) and the green one is the O-poisoned state where $\rho_o >0.9999$). The black region is the inactive phase. The horizontal white dashed lines indicate $\alpha = 1.0$ (RGG) and $p=6.0 \times 10^{-5}$ (ERN). The steps are $\Delta y = 0.005$ and $\Delta \alpha = 0.1$. In the ERN diagram there are 145 points in the vertical axis. The results for $y > 0.8$, $\alpha > 13$ and $p > 10^{-1}$ are not shown since there are no further relevant information on these regions.}
\label{phase_diagram}
\end{figure}

As shown above, the ZGB model considered in this work presents one active phase with production of CO$_2$ molecules, an O-poisoned state for low values of $y$ and $\alpha$ (RGG) or $p$ (ERN) and a CO-poisoned phase for large values of $y$. These three phases are also observed for the standard ZGB model on regular SL. In addition, the random networks produce another absorbing phase for low values of $\alpha$ and $p$ (see Fig. \ref{densidadesRGG}(a)) with the presence of both CO molecules, O atoms and vacant sites in a large number of components, however without production of CO$_2$ molecules. Here, for simplicity, we named this absorbing phase as the inactive phase. Figures \ref{phase_diagram}(a) and (b) show the phase diagram for the RGG and ERN, respectively, for $0 \leq y \leq 0.8$, $0 \leq \alpha \leq 13$, and $10^{-5} \leq p \leq 10^{-1}$. 
In the green colored region on both diagrams one finds the O-poisoned state and the red one represents the CO-poisoned state. The blue colored region shows the phase where there are sustainable production of CO$_2$ molecules, i.e., where the catalytic reactions occur. As can be seen on both phase diagrams, this active region occupies the majority of the area indicating that the CO$_2$ production is largely enhanced with the increase of $\alpha$ and $p$ (which, in turn, increases the average degree $\bar{k}$). Finally, the black colored region is the inactive region of the phase diagrams where the networks possess many components preventing the systems to percolate. For $\alpha < 1.0$ (for RGG) and $p < 5.0 \times 10^{-5}$ (for ERN), there are many components isolated from each other such that CO molecules can be trapped in one component, O atoms can be trapped in another one with two or more nodes, and there may be some vacant sites thus preventing the formation of any of the three phases present in the original model. For $\alpha=1.0$, $p=6.0 \times 10^{-5}$ (see the white dashed lines) and $0.4 < y < 0.6$, each phase diagram presents an appendix (intersection between the blue region and the white dashed line), i.e., a small region belonging to the active phase. As can be seen in Fig. \ref{topological}(a) and (b), these are the points where the networks start the processes of percolation even though there are many components yet. However, at these points the largest component is already large enough to start the catalytic reaction and the active phase is initiated. For increasing values of $\alpha$ and $p$ the other phases appear as well as the continuous and discontinuous phase transition points. 

Another important result we obtained in this study is that with increasing $\alpha$ and $p$, the continuous transition is eliminated for $y>0$ while the discontinuous transition continues to exist and stabilizes at $y \simeq 0.65$ for both random networks. Therefore, the ZGB model on random networks is capable to reproduce, at least qualitatively, the existence of only one phase transition, the discontinuous one, as predicted by experimental works \cite{golchet1978, matsushima1979, ehsasi1989, christmann1991, block1993}.

\subsection{Visualization of the network}

\begin{figure*}
\centering
\includegraphics[width=2.3 in]{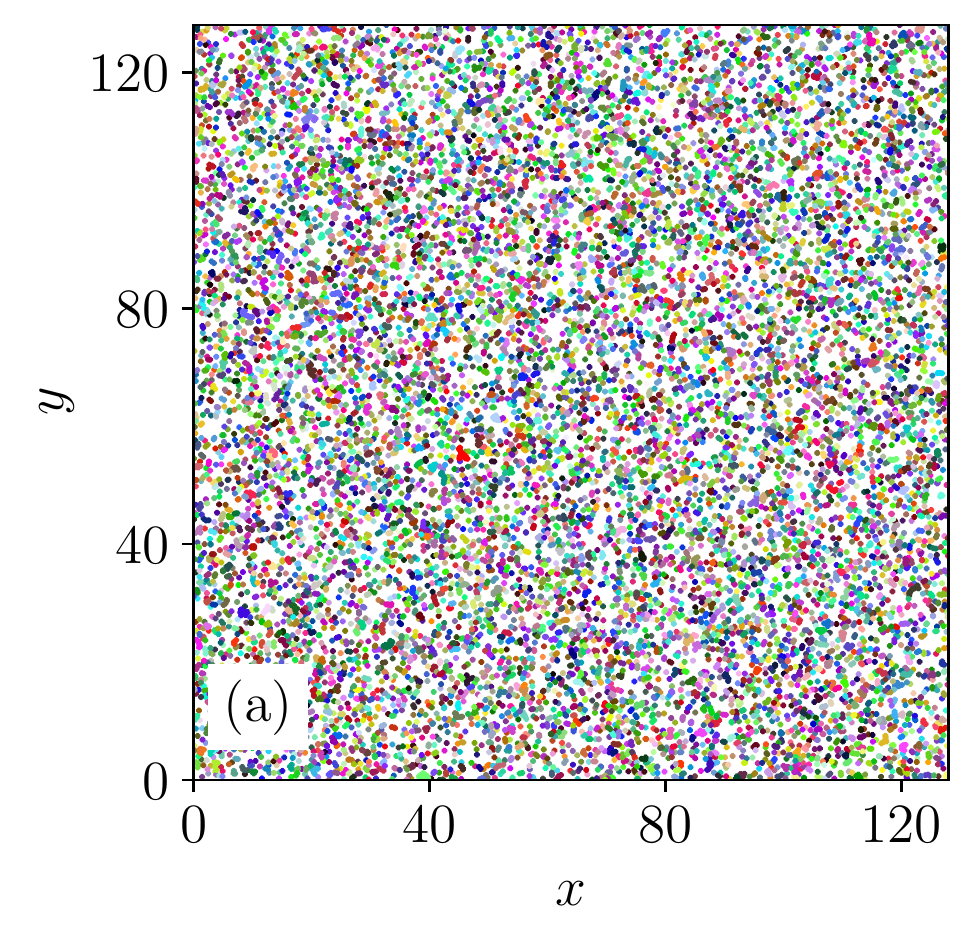}
\includegraphics[width=2.3 in]{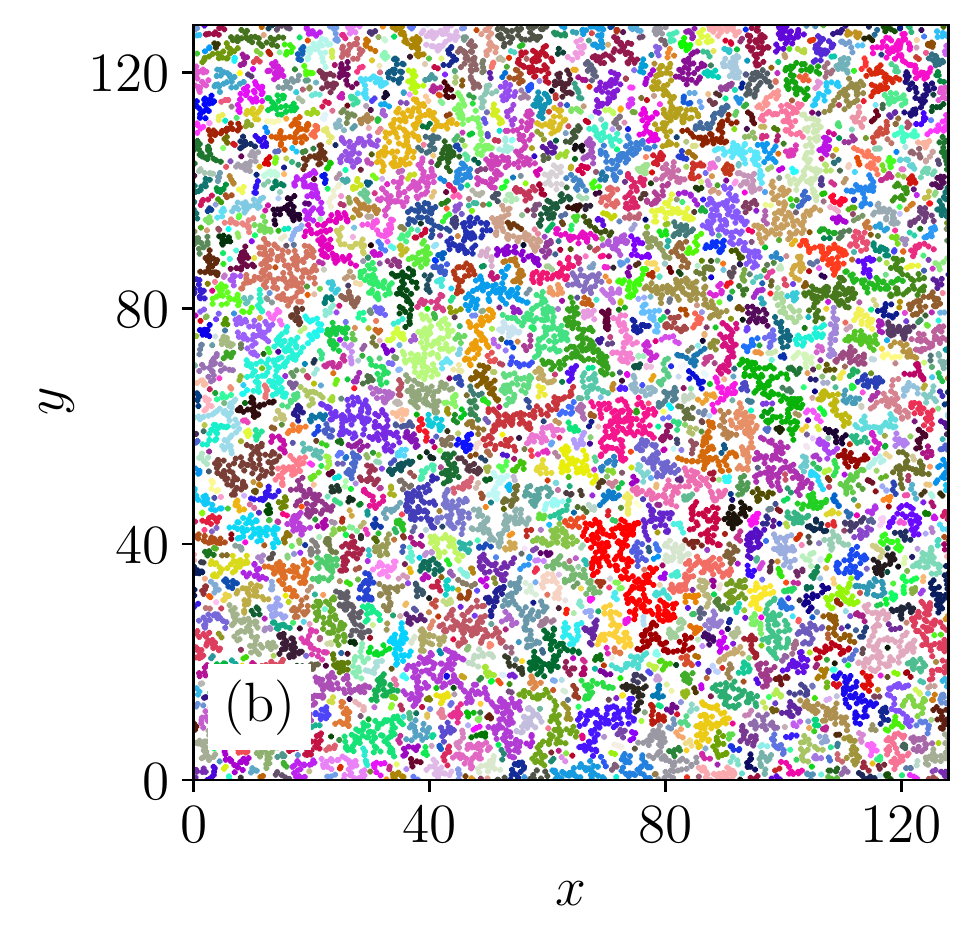}
\includegraphics[width=2.3 in]{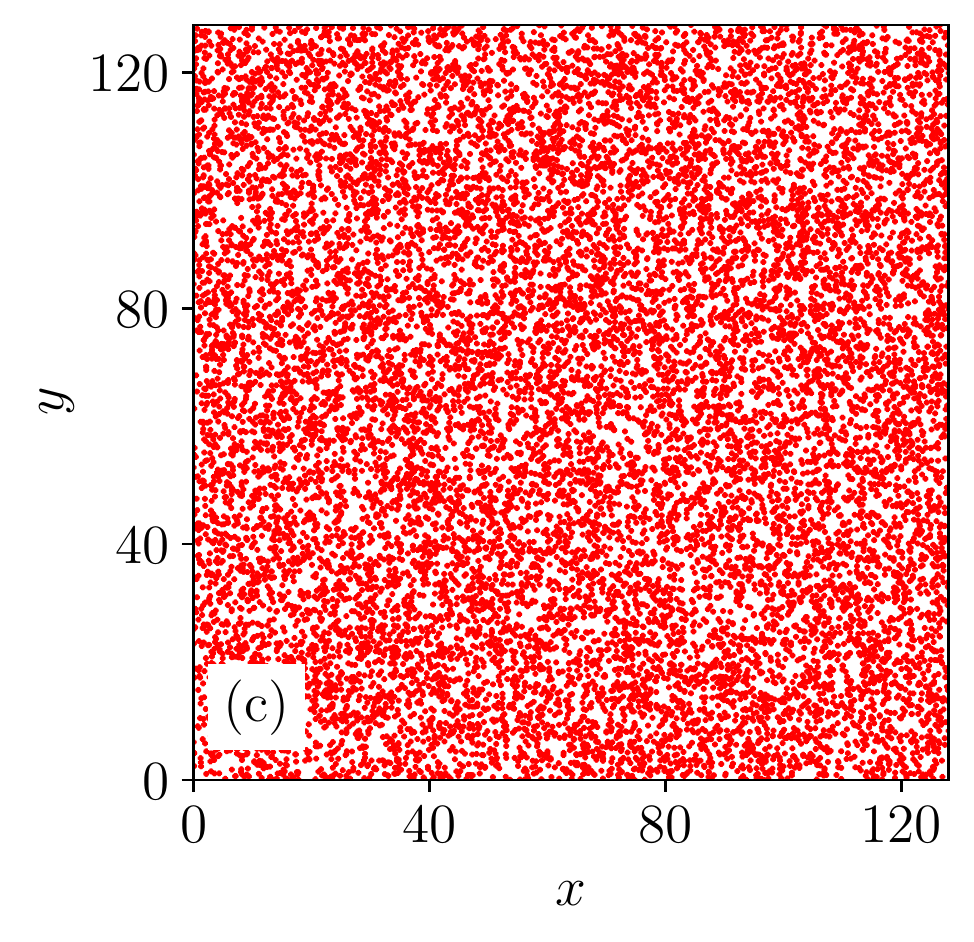}
\caption{The effect of the radius $\alpha$ on the topological properties of the RGG. The linear size of the square box is $L = \sqrt{M/\rho} = 128$. Each color represents a different component. The agents belonging to the largest component are painted in red. (a) $\alpha=0.5$, $\bar{k} \approx 0.79$, $n \approx 0.663$ and $s \approx 0.001$. (b) $\alpha=1.0$, $\bar{k} \approx 3.13$, $n \approx 0.131$ and $s \approx 0.013$. (c) $\alpha=2.0$, $\bar{k} \approx 12.58$, $n \approx 1/M$ and $s \approx 1.0$.} 
\label{RGG_comps}
\end{figure*}

In order to better understand and make an intuitive idea of the network topological properties, crucial to understand the phase diagrams, we present in this section snapshots of the graphs for different control parameters. To visualize the RGG and ERN for different values of $\alpha$ and $p$, respectively, we plotted a figure in which each component is represented by specific color. Figure \ref{RGG_comps} shows the snapshots of the RGG for three different values of $\alpha$ and the largest component is represented by the red dots.

In the Fig. \ref{RGG_comps}(a), we show the graph for $\alpha = 0.5$, which gives $\bar{k} \sim 0.79$ and creates a large number of components $\mathcal{N}=nM\simeq 0.663 \times 128^2\simeq 1.09\times 10^4$ (see Fig \ref{topological}). As the sites of different components are not connected, there can be O atoms isolated in some components, CO molecules isolated in another components, or even vacant sites, creating what we call the inactive phase. In the Fig. \ref{RGG_comps}(b), we show a network with $\alpha = 1.0$ which produces $n \sim 0.1$. At this point, the largest component (in red) starts to be visible and its size allows the production of CO$_2$ molecules for some values of $y$ (the intersection between the blue region and the white dashed line on the Fig. \ref{phase_diagram}(a)), although there is still no percolation. At $\alpha=2.0$ (Fig. \ref{RGG_comps}(c)), the network is connected (composed of only one component): $n \sim 1/M$ and $s \sim 1.0$, and the three phases of the original model can be observed.
 \begin{figure*}
 \centering
 \includegraphics[width=2.2 in]{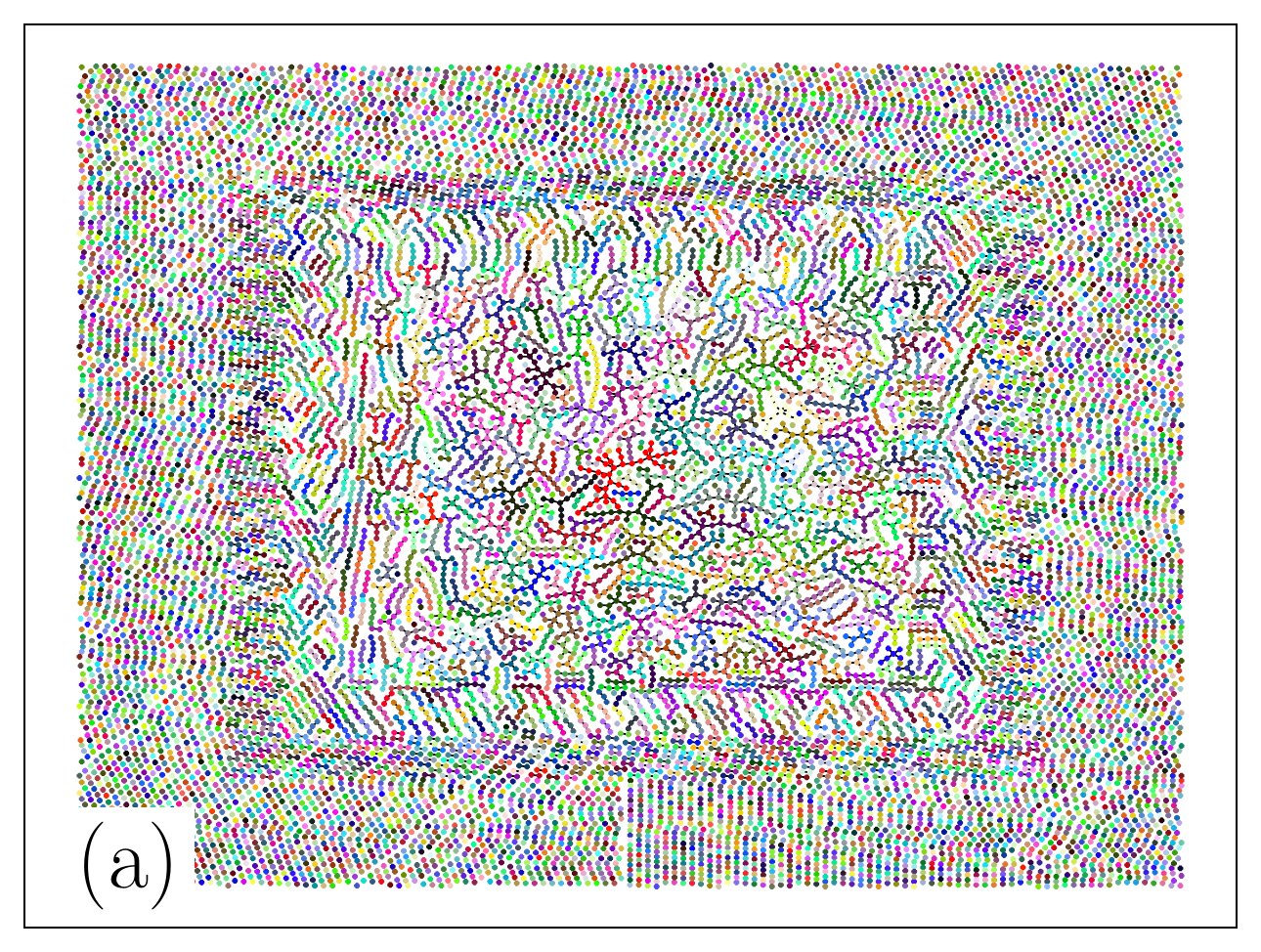}
 \includegraphics[width=2.2 in]{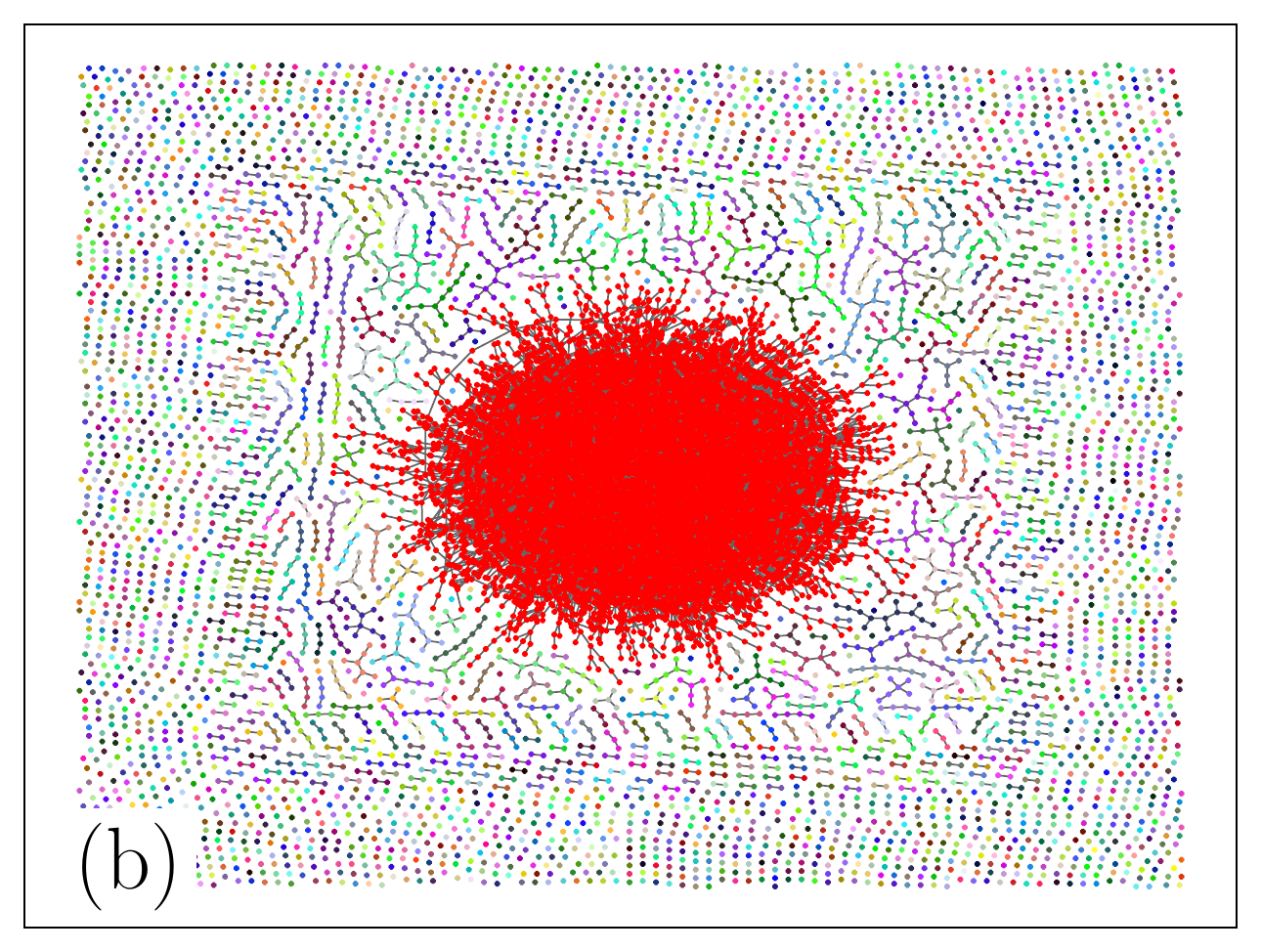}
 \includegraphics[width=2.2 in]{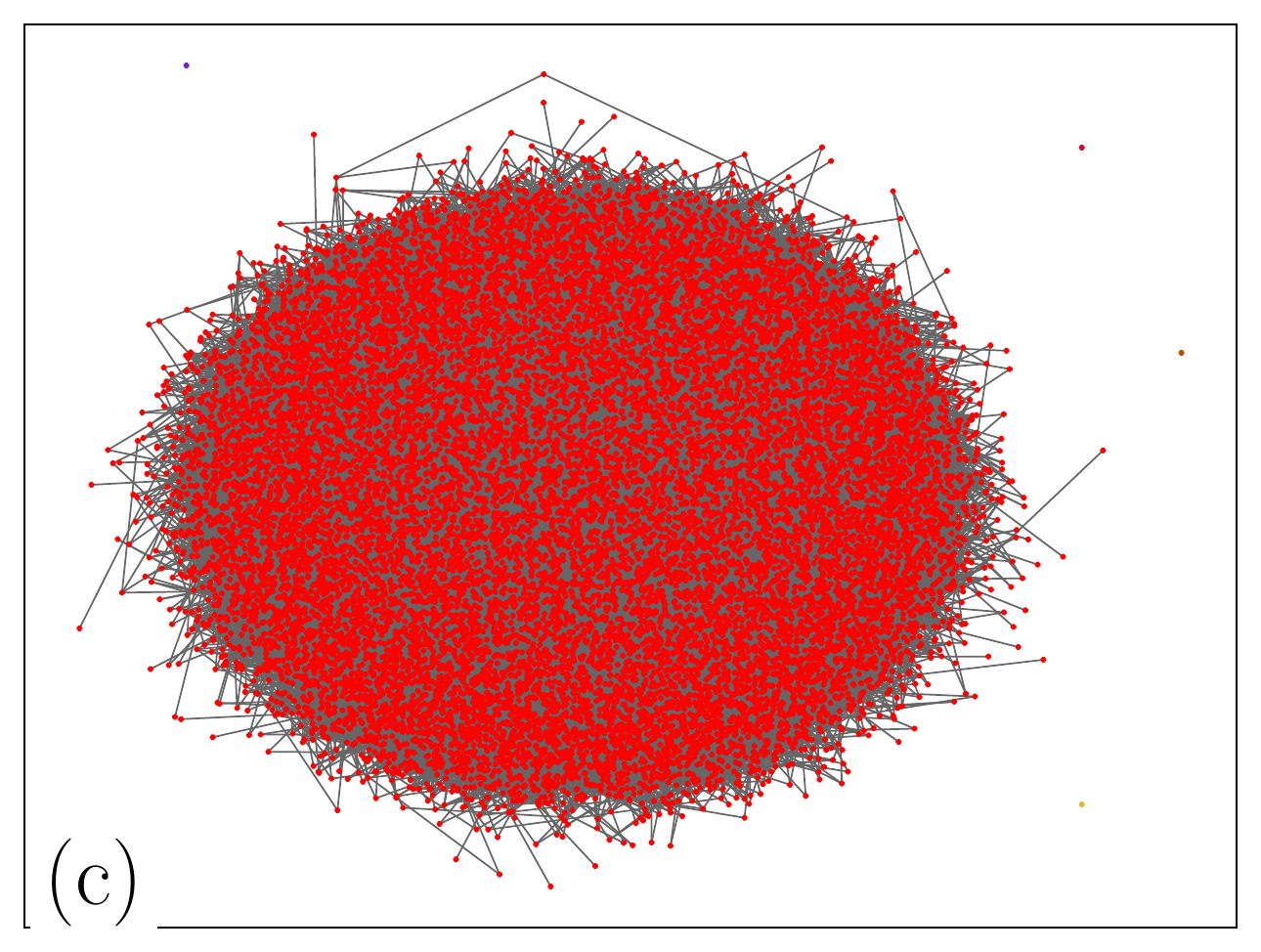}
 \caption{The effect of the connection probability $p$ on the topological properties of the ERN.  Each color represents a different component. The agents belonging to the largest component are painted in red. There is no spatial position concept in the ERN. These graphics were constructed in such a way that the largest component appears in the center and the isolated nodes in the outer border of the image. (a) $p= 5 \times 10^{-5}$, $\bar{k} \sim 0.8$, $ n \sim 0.59$ and $s \sim 0.005$. (b) $p=10^{-4}$, $\bar{k} \sim 1.6$, $ n \sim 0.245$ and $s \sim 0.661$. (c) $p= 5 \times 10^{-4}$, $\bar{k} \sim 8.2$, $ n \sim 1/M$ and $s \sim 1.0$.} \label{ERN_comps}
 \end{figure*}

Fig. \ref{ERN_comps} shows the snapshots of the ERN for three different values of $p$ and the largest component is also represented by the red dots. As the spatial position of the nodes has no meaning, we constructed the graphs in such a way that the isolated nodes are spread on the outermost part of the figure, the connected ones stay on the innermost part. The larger the degree closer to the center the node stays, so that the largest component is placed at the center of the box.

Although the ERN and RGG are conceived in a very different way, the analysis of both networks are similar. In the Fig. \ref{ERN_comps}(a), we plot a snapshot with $p=5\times 10^{-5}$ which produces $ n \sim 0.59$ such that most of the sites are isolated preventing the formation of active and poisoned phases (black region of Fig. \ref{phase_diagram}(b)). This behaviour is the expected one as the critical value for percolation is $p_c=1/M \sim 6.1 \times 10^{-5}$ as described above. For $p=10^{-4}$ (Fig. \ref{ERN_comps}(b)), the largest component grows up considerably ($s \sim 0.661$) as the system just started the percolation, enabling the appearance of the active phase and of a small appendix at the inactive phase of the Fig. \ref{phase_diagram}. The graph for $p= 5 \times 10^{-4}$ presented in Fig. \ref{ERN_comps}(c) shows that the percolated network is almost connected (a few sites are still isolated) and the active and poisoned phases appear

\section{Conclusions} \label{sec:conclusions}

In summary, we have performed steady-state Monte Carlo simulations in order to investigate the behavior of the phase transitions of the ZGB model when two different kinds of random networks are used as the surfaces where the catalytic reactions of the model occur. For these networks, the number of neighbors of a given site can be controlled leading to important changes in the phase diagram. For low values of $\bar{k}$ the system does not percolate and all three original phases of the ZGB model on the SL are absent. After the network percolates the system enters in the regime where all original ZGB three phases are present. A remarkable feature is the absence of the O-poisoning phase in both networks for large average degrees. This finding entails the absence of the continuous transition in agreement with experimental works.


Our work sets a new precedent by showing that the communication pattern among the sites on the catalytic surface plays an important role to the point of theoretically reproduce only the phase transition observed experimentally. Furthermore, the crucial properties of the network go far beyond the simple number of neighbors, as evidenced by the calculation of the volume of CO$_2$ molecules in the three evaluated networks. Models like ZGB and others, that have been traditionally implemented in the square lattice, can have their understanding improved when others networks are used. The interdisciplinary association of a catalytic surface model with the Network Science greatly enriches the mathematical tools used in the scope of Statistical Mechanics.

\begin{acknowledgments}

P.F. Gomes would like to thank S. M. Reia and J. F. Fontanari for helpful discussions. This work was partially supported by the Brazilian Agencies CNPq (H.A. Fernandes under the grant 408163/2018-6) and FAPEG (P.F. Gomes under the grant 17833 - 03/2015). H.A. Fernandes also would like to thank the financial support obtained from the Research and Innovation Coordination of the Federal University of Goi\'{a}s (COPI/UFG/REJ).

\end{acknowledgments}




\end{document}